\documentclass[acmsmall,screen]{acmart}

\AtBeginDocument{%
  \providecommand\BibTeX{{%
    \normalfont B\kern-0.5em{\scshape i\kern-0.25em b}\kern-0.8em\TeX}}}

\usepackage{fixltx2e}
\usepackage{amsmath, amsfonts}
\usepackage{graphicx}
\usepackage{algorithm}
\usepackage{algorithmicx}
\usepackage{algpseudocode}
\usepackage{amsmath,amsfonts}
\usepackage{array}
\usepackage{balance}
 \usepackage{booktabs}
\usepackage[skip=1pt,labelfont=bf]{caption}
 \usepackage{calligra}
\usepackage{color}
\usepackage{colortbl}
\usepackage{courier}
\usepackage{csvsimple}
\usepackage{enumitem}
\usepackage{fancybox}
\usepackage{fontenc}
\usepackage{graphicx}
\usepackage{listings}
\lstdefinelanguage{json}{
  morestring=[b]",% strings
  morecomment=[l]{//},% comments
  morecomment=[s]{/*}{*/},
  morekeywords={true,false,null},% keywords
  sensitive=false,
}

\usepackage{longtable}
\usepackage{lscape}
\usepackage{makecell}
\usepackage{marvosym}
\usepackage{moreverb}
\usepackage{multicol}
\usepackage{multirow}
\usepackage{pifont}% http://ctan.org/pkg/pifont%\usepackage{subfigure}
\usepackage{rotating}
\usepackage{setspace}
\usepackage{subcaption}
\usepackage[most]{tcolorbox}
\usepackage{threeparttable}
\usepackage{tikz}
\usepackage[normalem]{ulem}
\usepackage{url}
\usepackage{soul}
\usepackage{wasysym}
\usepackage{xspace}
\usepackage{xcolor}

\usepackage{url}
\usepackage{soul}
\usepackage{wasysym}
\usepackage{hyperref}   
\hypersetup{
  colorlinks=true,
  linkcolor=transparent,
  anchorcolor=transparent,
  citecolor=black,
  urlcolor=black
}

\definecolor{myred}{RGB}{200, 50, 50}

\definecolor{darkpastelred}{rgb}{0.76, 0.23, 0.13}
\definecolor{ao(english)}{rgb}{0.0, 0.5, 0.0}
\definecolor{darkpastelred}{rgb}{0.76, 0.23, 0.13}
\definecolor{ao(english)}{rgb}{0.0, 0.5, 0.0}
\definecolor{yellow}{RGB}{255,255,153}
\definecolor{grey}{RGB}{224,224,224}

\usepackage{subcaption}

\usepackage[nameinlink]{cleveref}

\usepackage{lineno}\usepackage{pifont}
\definecolor{myred}{RGB}{200, 50, 50}
\definecolor{darkpastelred}{rgb}{0.76, 0.23, 0.13}
\definecolor{ao(english)}{rgb}{0.0, 0.5, 0.0}

\definecolor{darkpastelred}{rgb}{0.76, 0.23, 0.13}
\definecolor{ao(english)}{rgb}{0.0, 0.5, 0.0}

\definecolor{yellow}{RGB}{255,255,153}
\definecolor{grey}{RGB}{224,224,224}

\usepackage{ifthen}

\newboolean{showcomments}
\setboolean{showcomments}{true} % true or false

\newcommand{\mynote}[2]{}

\ifthenelse{\boolean{showcomments}}{
  \renewcommand{\mynote}[2]{%
    \fbox{\bfseries\sffamily\scriptsize#1}%
    {\small$\blacktriangleright$\textsf{\emph{#2}}$\blacktriangleleft$}%
  }
}{}
\definecolor{DarkOrange}{rgb}{0.8,0.3,0.0}
\definecolor{DarkCyan}{rgb}{0.0, 0.55, 0.55}
\definecolor{DarkCyel}{rgb}{1.0, 0.49, 0.0}
\definecolor{yellow-green}{rgb}{0.6, 0.8, 0.2}

\newcolumntype{?}{!{\vrule width 1pt}}

\newcommand{\highlight}[1]{\begin{tcolorbox}[leftrule=0mm,rightrule=0mm,toprule=0mm,bottomrule=0mm,left=2pt,right=2pt,top=2pt,bottom=2pt]
  #1
  \end{tcolorbox}
}

\makeatletter
\DeclareRobustCommand\onedot{\futurelet\@let@token\@onedot}
\def\@onedot{\ifx\@let@token.\else.\null\fi\xspace}

\makeatother

\begin{document}

% \title{From Description to Fix: A Study of Textual and Semantic Relationships in Bug Resolution}
%\title{How Well Do Tests and Patches Reflect Bug Reports? A Large-Scale Alignment Study}

% \title{From Bug Reports to Fixes: A Large-Scale Study of Structural and Semantic Alignment Across Tests and Patches}
\title{Semantic Drift in Bug Resolution: How Behavioral Signals Propagate from Reports to Tests and Patches}

\author{Wendkûuni C. Ouédraogo}
\email{wendkuuni.ouedraogo@uni.lu}
\affiliation{
  \institution{University of Luxembourg}
 	\country{Luxembourg}
}

\author{Yinghua Li}\authornote{Corresponding author.}
\email{yinghua.li@njust.edu.cn}
\affiliation{
  \institution{Nanjing University of Science and Technology}
 	\country{China}
}

\author{Xueqi Dang}
\email{xueqi.dang@uni.lu}
\affiliation{
  \institution{University of Luxembourg}
 	\country{Luxembourg}
}

\author{Paweł Borsukiewicz}
\email{pawel.borsukiewicz@uni.lu}
\affiliation{
  \institution{University of Luxembourg}
  \country{Luxembourg}
}

\author{Liang Xiao}
\email{xiaoliang@mail.njust.edu.cn}
\affiliation{
  \institution{Nanjing University of Science and Technology}
 	\country{China}
}

\author{Lingfeng Bao}
\email{lingfengbao@zju.edu.cn}
\affiliation{
  \institution{Zhejiang University}
 	\country{China}
}

\author{Anil Koyuncu}
\email{anil.koyuncu@cs.bilkent.edu.tr}
\affiliation{
  \institution{Bilkent University}
 	\country{Turkey}
}

\author{Jacques Klein}
\email{jacques.klein@uni.lu}
\affiliation{
  \institution{University of Luxembourg}
  \country{Luxembourg}
}

\author{David Lo}
\email{davidlo@smu.edu.sg}
\affiliation{
  \institution{Singapore Management University}
  \country{Singapore}
}

\author{Tegawend\'e F. Bissyand\'e}
\email{tegawende.bissyande@uni.lu}
\affiliation{
  \institution{University of Luxembourg}
 	\country{Luxembourg}
}

\renewcommand{\shortauthors}{Ouédraogo et al.}

\begin{abstract}
Desc2Fix is a framework for measuring semantic alignment between bug reports (issues descriptions), triggering tests, and developer-written fixes. Alignment is operationalized through structured behavioral anchors (e.g., reproduction steps, API/exception cues, expected vs. actual behavior), deterministic similarity metrics (ROUGE, SBERT, CodeBERT, OpenAI embeddings), and LLM-based judgments grounded in coverage, correctness, and specificity. Our analysis covers 2,857 report–test–patch triplets from Defects4J and SWT-Bench using two widely adopted instruction-tuned LLMs from distinct model families (proprietary and open-weight). LLMs reliably extract structured signals (up to 90\% completeness) and exhibit strong cross-model consistency, yielding a stable semantic input contract for downstream reasoning. However, alignment is highly representation-sensitive: lexical similarity alone is insufficient; full diffs provide the most stable basis for judging report–patch correspondence; and structured summaries trade surface textual overlap for stronger correspondence at the level of individual actions and entities. Across more than 182,000 LLM-based alignment judgments, Desc2Fix reveals that both models exhibit systematic optimism relative to humans (1–2 points on 5-point scales) and only modest rank agreement, motivating bias-aware evaluation. Our results demonstrate that behavioral alignment is measurable but not reducible to similarity, and that structured anchors combined with embedding-based proxies provide reproducible signals for ranking and filtering tests and candidate patches. By transforming alignment into a controllable engineering signal, Desc2Fix lays the groundwork for more reliable test generation, semantics-aware fault localization, principled patch ranking, and improved bug report authoring.

\end{abstract}




\maketitle

\section{Introduction}
\label{sec:intro}

Software systems are inherently prone to defects, making bug resolution a central task in software maintenance. In modern development workflows, bug reports are the primary interface for communicating software anomalies~\cite{bissyande2013got}. They typically describe reproduction steps, input conditions, stack traces, and expected versus actual behavior~\cite{lamkanfi2011comparing,zhou2012should}, providing the semantic basis for constructing triggering tests and corrective patches. Yet, these behavioral signals are often diluted or partially lost as they propagate to executable artifacts. Tests may reproduce symptoms without encoding intended behavior, and patches may suppress failures without restoring semantic correctness. This \emph{semantic drift}~\cite{rastkar2010summarizing} weakens the correspondence between report, test, and fix, limiting precision, interpretability, and automation reliability. Bridging this gap between natural-language intent and code-level realization remains a fundamental challenge in automated software maintenance~\cite{koyuncu2019ifixr,gao2015fixing}.

Recent advances in automated testing, fault localization, and program repair have substantially improved maintenance efficiency~\cite{zhou2012should,liu2013r2fix,wang2014version,youm2015bug,koyuncu2019ifixr}. However, these techniques typically treat bug reports as textual inputs rather than as structured semantic drivers, often overlooking contextual anchors such as input conditions, exception types, or expected outcomes. At the same time, large language models (LLMs) and code-aware transformers have enabled joint reasoning over natural language and code~\cite{chen2021evaluating,roziere2023code,jin2023inferfix,zhang2404autocoderover}. Fault localization and repair systems have progressed from lexical retrieval to semantic and agentic reasoning~\cite{zhou2012should,zhang2019finelocator,fang2021classification,chakraborty2024rlocator}, and LLM-based test generation can now reproduce failures with increasing accuracy~\cite{plein2024automatic,kang2023large,feng2024prompting}. Yet these stages are still evaluated largely in isolation. Recent work has begun to approach this gap from adjacent angles, whether by refining bug report specifications with repository-derived evidence for repair agents~\cite{fahim2026bug} or by identifying which report-level features matter for agent-based resolution~\cite{khatib2026makes}, but neither directly measures how behavioral signals are preserved or transformed across the report-test-patch pipeline.

A fundamental question thus remains largely unexplored: \textbf{How are the behavioral signals expressed in bug reports preserved, transformed, or lost as they propagate to triggering tests and corrective patches?} Answering this question is critical not only for understanding the interpretability of current LLM-based systems, but also for designing maintenance pipelines where bug reports act as controllable semantic drivers rather than passive textual inputs. Quantifying cross-artifact alignment could provide an actionable engineering signal: reports that expose strong behavioral anchors can be transformed into reliable triggering tests and precise repair objectives, whereas weakly aligned tests may validate superficial fixes rather than semantic correctness, and patches that diverge from report-level intent may optimize for fail-to-pass transitions without restoring intended functionality. Measuring alignment could therefore enable (i)~early detection of non-actionable reports, (ii)~alignment-guided test generation and filtering, (iii)~semantics-aware fault localization, and (iv)~principled ranking of candidate patches in automated repair workflows.

To operationalize this perspective, we introduce \textsc{Desc2Fix}, a unified framework for quantifying cross-artifact semantic alignment. We first extract structured behavioral anchors (e.g., reproduction steps, API/exception cues, expected vs.\ actual behavior) using LLM-assisted signal extraction, producing a controlled representation of developer intent. We then define alignment dimensions capturing how entity-level, scenario-level, and behavioral signals are reflected in tests and patches. Alignment is assessed through complementary perspectives: deterministic similarity metrics (lexical and embedding-based), LLM-based semantic judgments grounded in coverage, correctness, and specificity, and human annotations used to validate model outputs. By triangulating these signals, \textsc{Desc2Fix} enables fine-grained analysis of semantic preservation and drift across the bug resolution pipeline.

Our empirical evaluation spans two complementary datasets, complementary along both language and curation dimensions, enabling cross-language and cross-artifact analysis. Defects4J v3.0.1~\cite{just2014defects4j} provides curated Java bugs with explicitly linked reports, triggering tests, and patches, supporting controlled alignment analysis. SWT-Bench~\cite{mundler2024swt} extends the study to Python and real-world, less curated GitHub issues paired with validated fail-to-pass tests, capturing more diverse development settings. Together, these corpora allow us to examine how behavioral signals propagate across artifacts and ecosystems, making this the first large-scale cross-language study of semantic alignment among bug reports, triggering tests, and corrective patches.

\noindent\textbf{This paper makes the following contributions:}
\begin{itemize}[leftmargin=*]

\item \textbf{A unified framework for cross-artifact behavioral alignment.}
We introduce \textsc{Desc2Fix}, a principled framework for quantifying how behavioral signals propagate from bug reports to triggering tests and corrective patches. The framework combines LLM-assisted structured extraction, a practical characterization of alignment dimensions (Entity, Scenario, Behavioral) grounded in established bug-report structure, and multi-perspective evaluation (lexical, embedding-based, and LLM-based), enabling fine-grained and human-grounded analysis.

\item \textbf{A large-scale cross-language empirical study of semantic propagation.}
Across Defects4J (Java) and SWT-Bench (Python), we analyze 2,857 report--test--patch triplets to characterize how behavioral signals are preserved, transformed, or lost. Our study reveals representation-sensitive alignment effects, systematic LLM optimism, and the limitations of similarity-based proxies.

\item \textbf{\textsc{Desc2Fix}: a reusable alignment benchmark.}
We release an annotated corpus linking structured bug reports, tests, and patches, enriched with embedding-based similarity scores and LLM-evaluated alignment judgments, supporting reproducible research on semantics-aware testing and repair.

\item \textbf{Actionable guidance for semantics-aware maintenance.}
We identify which report attributes (e.g., explicit expected/actual contrast, salient steps) most strongly influence how faithfully a test encodes reported behavior and how precisely a patch addresses it, and we sketch alignment-guided engineering blueprints, operationalized as reference algorithms, for test generation and repair workflows.

\item \textbf{A fully reproducible research package.}
All datasets, prompts, evaluation scripts, and analysis code are publicly released\footnote{\url{https://anonymous.4open.science/r/Desc2Fix-EC04/}} to facilitate replication and extension in LLM-based maintenance research.

\end{itemize}

The paper is organized as follows: \Cref{background} outlines key concepts. \Cref{approach} details our study design. \Cref{results} presents findings for the three research questions. \Cref{discussion} discusses implications and limitations. \Cref{relatedwork} positions our study within existing literature, and \Cref{conclusion} concludes with key insights and future directions.

\section{Background}
\label{background}

%“quotes” 
\subsection{Bug Reports as Semantically Rich Artifacts}
\label{}
Bug reports\footnote{Throughout this paper, we use the terms bug report and issue interchangeably, following recent usage in the agentic and LLM-based software engineering literature (e.g., SWE-bench and its derivatives), where the same natural-language artifact reporting a defect is variably referred to as a “bug report” or an “issue.”} are the primary entry point of the bug resolution process, describing failure context, reproduction scenarios, and expected versus actual behaviors~\cite{bissyande2013got,lamkanfi2011comparing}. Beyond documenting symptoms, they encode linguistic and semantic signals that guide downstream activities such as test creation, fault localization, and patch generation. Report quality strongly influences resolution effectiveness: detailed reproduction steps and precise behavioral 
descriptions accelerate fixes~\cite{bettenburg2008makes}, and structural and linguistic features correlate with fix success~\cite{nguyen2012multi}. However, bug reports exhibit substantial variability in verbosity, terminology, and contextual explicitness, which can cause semantic drift when transitioning to tests and patches — key failure cues may be paraphrased, transformed, or omitted entirely.

\subsection{From Fault Localization to Automated Program Repair}
\label{}
Fault localization aims to connect natural-language bug reports to faulty code elements. Early information-retrieval approaches (e.g., BugLocator~\cite{zhou2012should}, AmaLgam~\cite{wang2014version}, D\&C~\cite{koyuncu2019d}) relied on lexical similarity between reports and source files, remaining sensitive to vocabulary variation and contextual ambiguity~\cite{lee2018bench4bl}. Later techniques incorporated embedding-based models and neural classifiers~\cite{zhang2019finelocator,fang2021classification}, and more recently LLM-assisted methods further improved localization by reformulating reports or optimizing ranking objectives~\cite{shao2024enhancing,chakraborty2024rlocator}.
Automated Program Repair (APR) extends this bridge from textual intent to executable fixes. Early text-driven systems demonstrated the feasibility of leveraging bug reports for patch generation~\cite{liu2013r2fix,gao2015fixing}, including iFixR~\cite{koyuncu2019ifixr}, which drives repair by combining IR-based fault localization with fix-pattern-based patch generation directly from bug reports, while recent retrieval-augmented and agentic LLM systems synthesize validated patches directly from repository context~\cite{jin2023inferfix,zhang2404autocoderover,motwani2023better}. Despite this progress, most approaches optimize individual stages without explicitly examining semantic continuity across artifacts, the gap our study directly addresses.

\subsection{Test Generation from Bug Reports}
\label{}
Test generation translates bug reports into executable specifications that reproduce observed failures. Early code-centric tools such as EvoSuite~\cite{fraser2011evosuite} and Randoop~\cite{pacheco2007feedback} largely ignored contextual information embedded in bug reports, limiting their ability to capture failure-specific inputs and behavioral constraints. More recent LLM-based approaches generate 
tests directly from issue descriptions, though a gap persists between syntactic executability and true behavioral relevance~\cite{plein2024automatic}. Extracting explicit and implicit inputs from reports has been shown to substantially improve test 
reproducibility~\cite{ouedraogo2025enriching,ouedraogo2024extracting}, and large-scale evaluations confirm the potential of LLM-driven test synthesis across diverse 
benchmarks~\cite{kang2023large,feng2024prompting,al2025llput,ahmed2025otter,mundler2024swt}. However, existing studies primarily assess test correctness or executability rather than how behavioral cues propagate from reports into tests and subsequently into patches.

\subsection{The Need for Cross-Artifact Semantic Alignment}
\label{}
Fault localization, test generation, and program repair have each advanced significantly. While some recent systems already combine multiple stages, notably IR-based fault localization with patch generation~\cite{koyuncu2019ifixr}, or fault localization, edit, and validation within a single agentic loop~\cite{zhang2404autocoderover,bouzenia2025repairagent}, most of this integration remains outcome-driven: stages are chained to produce a working patch, without explicitly verifying that the behavioral intent expressed in the originating bug report is preserved at each step. Three limitations persist across this body of work. First, most approaches rely on lexical similarity or task-specific optimization~\cite{zhou2012should,wang2014version}, inadequately capturing deeper correspondences between natural-language intent and executable behavior. Second, no unified framework systematically quantifies how behavioral signals propagate across artifacts: prior work combining report- and test-based signals improves repair performance~\cite{koyuncu2019ifixr,motwani2023better} but does not measure how cues are preserved, transformed, or lost from description to test to patch. Third, existing evaluations target generation performance within individual benchmarks~\cite{just2014defects4j,mundler2024swt,jimenez2023swe} rather than cross-artifact semantic alignment. It is precisely this gap, measuring semantic preservation across the full report-test-patch pipeline rather than optimizing any single stage in isolation, that \textsc{Desc2Fix} is designed to close, through a large-scale, cross-language examination of alignment among bug reports, tests, and patches.
\section{Study design}
\label{approach}

\subsection{Analysis Overview}
\label{pipeline_overview}

\begin{figure*}[ht]
    \centering
    \includegraphics[width=\textwidth,scale=3]{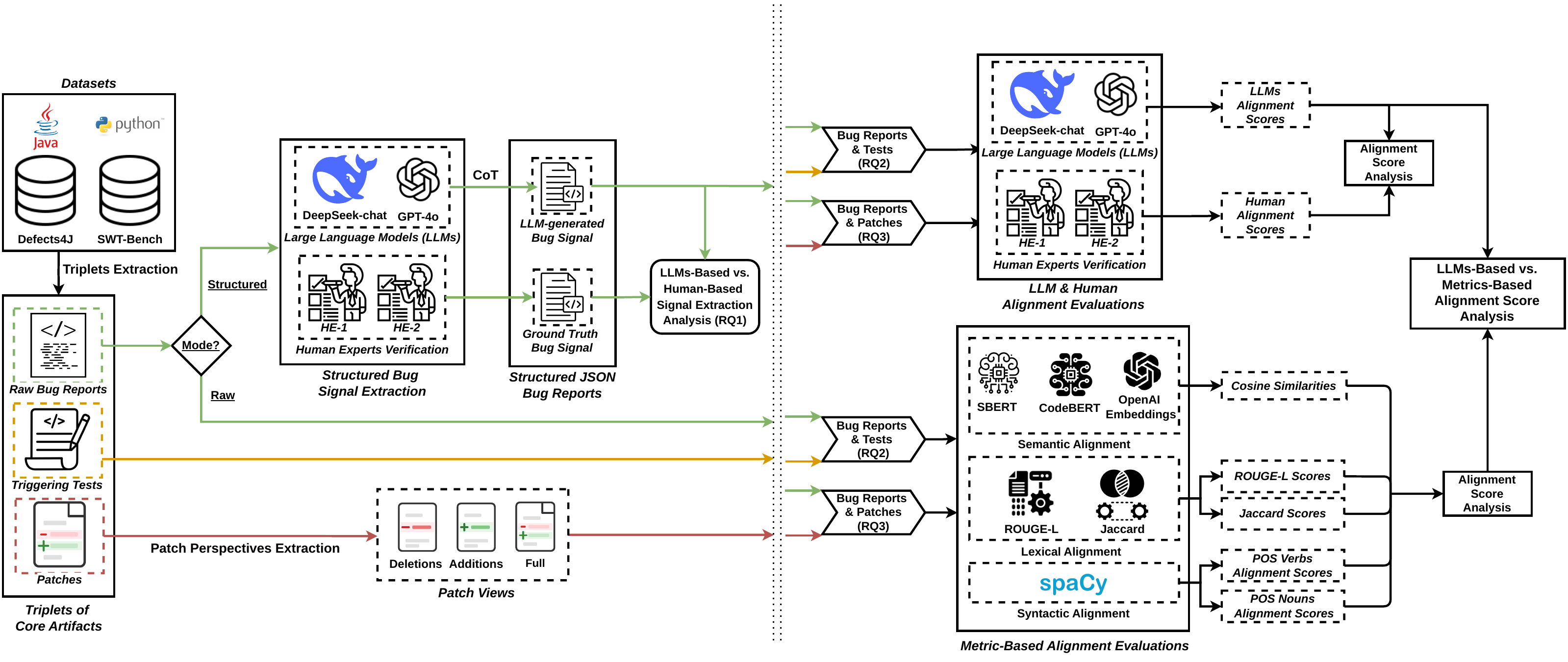}
    \caption{Overview of the general workflow of Desc2Fix.}
    \label{subsec:approach-overview}
\end{figure*}

Our methodology (Figure~\ref{subsec:approach-overview}) is designed to analyze the textual and semantic alignment among three core artifacts involved in software bug resolution: the \emph{bug report}, the \emph{triggering test}, and the \emph{patch}. We define a multi-layered analysis framework based on signal extraction, alignment taxonomy, and both metric- and LLM-based evaluation strategies.

%***************************************************************
\subsection{Textual and Semantic Signal Extraction}
\label{sub:signal_extraction}

We extract high-level semantic signals from bug reports to support alignment analyses with triggering tests and corrective patches. In our setting, bug reports are treated as \emph{pre-fix problem descriptions}, i.e., the developer-facing specification available \emph{before} (or at the time) a failing test and a corrective patch are produced. This distinction matters because post-fix narratives (e.g., retrospective summaries) may directly incorporate solution details and would inflate apparent cross-artifact alignment. We note, however, that the triggering test itself is not necessarily temporally independent of this post-fix knowledge: in practice, developers frequently author or finalize the triggering test alongside the patch, once the root cause is understood, rather than strictly before it. Report-test alignment (RQ2) should therefore be interpreted as measuring how well the test ultimately committed to the repository encodes the reported failure, not as evidence that the test was authored purely from the report in isolation; we revisit this timeline ambiguity in our discussion of report completeness (Section~\ref{subsec:completeness}). Signals are extracted at three abstraction levels:

\begin{itemize}
    \item \textbf{Entities}: exception types, API methods, constants, and error messages.
    \item \textbf{Scenarios}: reproduction steps and input conditions required to trigger the bug.
    \item \textbf{Behaviors}: expected vs.\ actual outcomes, including inferred fix intent.
\end{itemize}

We prompt the LLM to produce a structured JSON representation aligned with these levels, improving consistency in the presence of paraphrases or implicit cues.\footnote{\url{https://anonymous.4open.science/r/Desc2Fix-EC04/}} An example output is shown below:

\vspace{0.5em}
{\scriptsize
\begin{tcolorbox}[colback=gray!3, colframe=gray!40, title=\textbf{Example: Structured Extraction Output (JSON)}, left=4pt, right=4pt, top=4pt, bottom=4pt, boxrule=0.4pt, breakable]
\begin{lstlisting}[language=json]
{
  "exception": "NullPointerException",
  "api_involved": ["Form.submit"],
  "input_conditions": ["form is empty"],
  "expected": "Validation error is shown",
  "actual": "Application crashes with NullPointerException",
  "reproduction_steps": ["open the form",
                         "leave all fields blank",
                         "click submit"]
}
\end{lstlisting}
\end{tcolorbox}
}
\vspace{0.5em}

This representation enables systematic comparison across artifacts (e.g., whether reported exceptions, input conditions, and behavioral constraints are reflected in tests or patches), supporting our analysis of which signals are preserved, transformed, or lost throughout bug resolution. Our extraction schema is related to, but distinct from, prior structured bug-report representations. BEE~\cite{song2020bee} classifies reports into Observed Behavior, Expected Behavior, and Steps to Reproduce using a machine-learning classifier, and ChatBR~\cite{bo2024chatbr} uses ChatGPT to detect and generate this same information when missing; both operate on the bug report in isolation, aiming to improve or complete the report itself rather than to compare it against downstream artifacts. Similarly, multi-dimensional LLM-based assessment of testing reports~\cite{wang2025multi} scores a single artifact's quality along several axes but does not measure correspondence across artifacts. In contrast, our six-field schema (exception, API, input conditions, reproduction steps, expected, actual) is extracted specifically to serve as a comparison basis against triggering tests and patches, and is evaluated not for report quality in isolation but for how well its signals propagate to, or are preserved in, downstream executable artifacts.

%***************************************************************
%\subsection{Taxonomy of Alignment Dimensions and Metrics}
\subsection{Practical Characterization of Alignment Dimensions and Metrics}
\label{sub:alignment_dimensions}

To analyze how semantic signals propagate from bug reports to tests and patches, we characterize three alignment dimensions, each evaluated using lexical overlap and embedding-based semantic similarity. \textbf{Entity Alignment} measures whether report-level entities (e.g., exception types, APIs, literals) appear in corresponding tests or patches, building on prior evidence that entity-level cues (exception types, API names) are among the most consistently reused signals when reports are translated into code-level artifacts~\cite{zhou2018recognizing,zhou2023leveraging}. \textbf{Scenario Alignment} assesses whether the reproduction steps and input conditions that describe \emph{how} to trigger the bug, i.e., the concrete sequence of actions and preconditions reported by the user or developer, are reflected in test setup or execution structure; this dimension follows prior work showing that explicit reproduction steps are among the report elements most strongly associated with developer usefulness and fix success~\cite{bettenburg2008makes,nguyen2012multi}. \textbf{Behavioral Alignment} captures whether the expected and actual outcomes reported are enforced in test assertions or addressed in patch logic, consistent with the well-established view that the expected-versus-actual contrast is a central, high-value component of bug report quality~\cite{bettenburg2008makes,zimmermann2010makes}. For lexical alignment, we use literal and fuzzy overlap metrics (Jaccard, ROUGE-L) and POS-based verb matching for scenarios. For semantic alignment, we compute similarity using SBERT, CodeBERT, BERTScore (behavioral
dimension), and OpenAI embeddings. Each bug report is structured via LLM extraction (Section~\ref{sub:signal_extraction}); alignment scores are then computed per dimension to analyze signal preservation, transformation, and loss across artifacts.

% %***************************************************************
\subsection{LLM-Assisted Semantic Analysis}
\label{sub:llm_analysis}

To complement metric-based analyses, we use large language models (LLMs) as semantic evaluators of alignment between bug reports and corresponding tests or patches. We consider two report representations: \textbf{RAW}, where the model receives the original bug report in free-form text, and \textbf{STRUCTURED}, where it is provided with the extracted JSON representation described in Section~\ref{sub:signal_extraction}. This design allows us to assess the impact of the STRUCTURED representation on semantic evaluation.
For each report--artifact pair, the LLM produces a structured JSON assessment with three Likert-scale scores (1--5): \emph{coverage}, \emph{correctness}, and \emph{specificity}, each accompanied by a brief justification. A final alignment score is computed as the rounded average of the three dimensions. This protocol yields a standardized, machine-parseable output format across models and report representations, which we use to systematically compare the RAW and STRUCTURED representations and to quantify agreement with human annotators (Sections~\ref{subsec:RQ2}--\ref{subsec:RQ3}); it does not by itself guarantee that the resulting scores are consistent or well-calibrated, a question we examine empirically in our results.
We evaluate two widely adopted instruction-tuned LLMs representing distinct model families: \textbf{GPT-4o}~\cite{hurst2024gpt}, a state-of-the-art commercial model, and \textbf{DeepSeek-Chat}~\cite{liu2024deepseek}, a competitive open-weight alternative. Both are accessed via API under consistent decoding settings (temperature$=1.0$, top\_p$=1.0$), and all responses are serialized in JSON for downstream aggregation. Prompt templates and evaluation details are available in the replication package.
% % %***************************************************************
\subsection{Research Questions}
\label{sub:RQ}

Our study investigates whether cross-artifact alignment can serve as a measurable and actionable signal for improving debugging and repair pipelines.

\vspace{0.2cm}
\noindent\textbf{RQ1: Can LLMs reliably extract structured semantic signals from bug reports to enable downstream alignment analysis?}  
Before alignment can be quantified, bug reports must be transformed into structured representations capturing actionable anchors (e.g., APIs, input conditions, reproduction steps, expected/actual behavior). We evaluate GPT-4o and DeepSeek-Chat as independent extractors and assess whether their outputs are sufficiently complete, coherent, and human-aligned to serve as reliable inputs for repair-oriented analyses.

\vspace{0.2cm}
\noindent\textbf{RQ2: Does report--test alignment serve as a signal of how faithfully a triggering test
encodes the reported behavior, and which report properties make that encoding easier?}
We quantify the semantic and structural alignment between bug reports and triggering tests, and analyze which report characteristics (e.g., explicit expected/actual statements, API mentions, reproduction steps) are associated with stronger alignment. Rather than claiming that alignment guarantees test adequacy in the fault-detection sense, we examine whether alignment signals can identify tests that faithfully encode the reported failure and reports that are more readily translated into such tests.

\vspace{0.2cm}
\noindent\textbf{RQ3: Can report--patch alignment act as a semantic validation signal for program repair?}  
We measure how closely patches address the behaviors and conditions described in bug reports across multiple patch views (full diff, additions, removals). We analyze whether alignment scores and structured anchors can help distinguish semantically grounded fixes from potentially overfitting or incomplete patches.

% % %***************************************************************
\subsection{Prompting Techniques}
\label{sub:prompting_techniques}

All LLM-based evaluations use standardized prompts following the same structure: contextual grounding, explicit evaluation criteria, and JSON-formatted outputs. We adopt a lightweight Chain-of-Thought strategy~\cite{wei2022chain}, instructing the model to briefly reason before producing structured scores. Each task is evaluated under two variants: \textbf{RAW} (natural-language report) and \textbf{STRUCTURED} (JSON-based representation from Section~\ref{sub:signal_extraction}). Prompts are executed with GPT-4o and DeepSeek-Chat (temperature=1.0, top\_p=1.0). Full templates are available in the replication package.

\paragraph{Bug Report--Test Alignment}
For report--test pairs, the model evaluates alignment along four dimensions: coverage, correctness, specificity, and overall alignment, providing a brief justification followed by a structured JSON assessment with scores (1--5):

{\scriptsize
\begin{tcolorbox}[colback=gray!5, colframe=gray!40, 
title=\textbf{Bug Report--Test Alignment Output Schema}, 
left=2pt, right=2pt, top=2pt, bottom=2pt, boxrule=0.4pt, breakable]
\begin{lstlisting}[language=json]
{
  "alignment_score": <1-5>,
  "coverage": <1-5>,
  "correctness": <1-5>,
  "specificity": <1-5>,
  "justification": "..."
}
\end{lstlisting}
\end{tcolorbox}
}

\paragraph{Bug Report--Patch Alignment}
Patch evaluation follows the same scoring scheme. The model receives either the RAW or STRUCTURED report together with the patch (full diff, additions only, or deletions only) and produces scores using the same JSON schema above.

\paragraph{On model and prompting choices.}
Our experimental design spans two datasets, six conditions combining the two report representations with the three patch views, and four scoring dimensions. This generates a large volume of alignment ratings (over 182,000 in total), which imposes practical constraints on model choice and prompting complexity. We rely on Chain-of-Thought prompting rather than native reasoning (``thinking'') modes or agentic orchestration for two reasons: first, native reasoning and agentic modes substantially increase per-call cost and latency, which is prohibitive at this scale; second, and more fundamentally, our objective is to assess behavioral alignment as a \emph{measurable and reproducible signal} derived from a fixed, uniform prompting protocol, rather than to maximize generation or reasoning performance, which agentic orchestration is designed to optimize but at the cost of run-to-run variability that would undermine the comparability of scores across datasets, representations, and patch views. The two selected models represent complementary paradigms, proprietary versus open-weight, ensuring that our findings are not artifacts of a single provider's design choices. This uniform, JSON-structured prompting protocol also enables automated parsing and direct comparison of LLM-based scores with metric-based measures throughout our large-scale alignment analysis. We further discuss the implications of model recency and prompting strategy in our threats to validity (Section~\ref{subsec:threats-to-validity}).

\subsection{Datasets and Experimental Scope}
\label{sub:datasets}
We evaluate our framework on two complementary benchmarks covering Java and Python ecosystems: \textbf{Defects4J} and \textbf{SWT-Bench}. This dual-dataset design enables cross-language analysis of semantic alignment across bug reports, tests, and patches.
We curate a structured corpus from Defects4J v3.0.1~\cite{just2014defects4j}, focusing on 16 projects with accessible and consistent issue metadata (excluding JFreeChart). For each bug, we reconstruct triplets linking the natural-language report, triggering/relevant tests, and the human-written patch. Bug reports are retrieved from JIRA, GitHub, or archived sources. Patches are parsed into three \emph{patch views}: the full diff (both added and removed lines, as committed), additions only (only the lines the patch introduces, with removed lines excluded entirely rather than merely repositioned as in a split-diff display), and deletions only (only the lines the patch removes). These single-sided views isolate the marginal contribution of what a patch introduces versus what it discards to judged alignment (RQ3, Section~\ref{subsec:RQ3}). Tests are linked using \texttt{trigger\_tests/} and \texttt{relevant\_tests/} metadata: a \emph{trigger test} fails before the patch and passes after it, providing direct fail-to-pass evidence of the bug, whereas a \emph{relevant test} is any test linked to the fix commit, including regression tests that already passed beforehand; unless otherwise specified, \emph{triggering test} refers to the former throughout this paper. This process yields 511 curated triplets spanning 16 Java projects. Table~\ref{tab:defects4j_stats} summarizes corpus statistics.

\begin{table}[ht]
\vspace{0.5em}
\centering
\caption{Statistics of the curated \textsc{Defects4J-v3} corpus.}
\label{tab:defects4j_stats}
\scalebox{0.7}{
\begin{tabular}{@{}p{3.2cm} p{6.4cm} p{9.2cm}@{}}
\toprule
\textbf{Metric} & \textbf{Value} & \textbf{Interpretation} \\
\midrule
Total instances (RAW) & 511 & Triplets (bug report--test--patch) reconstructed from Defects4J v3. \\
Unique projects & 16 & Java projects spanning multiple domains. \\
Bug reports (retrieved) & 511 (100\%) & Available via JIRA, GitHub, or Google Code. \\
Avg.\ bug report length & 142.7 words & Textual richness of issue descriptions. \\
Avg.\ patch size & 52.4 lines & Overall modification granularity (src + test). \\
Avg.\ source patch size & 38.1 lines & Code fix magnitude. \\
Avg.\ test patch size & 14.3 lines & Extent of test modification. \\
Bugs with test patch & 383 (74.9\%) & Presence of test code in the correction. \\
Avg.\ trigger tests & 1.2 per bug & Tests failing on the buggy version. \\
Avg.\ relevant tests & 2.5 per bug & Tests explicitly linked to the bug. \\
Add/remove ratio & 1.31 & Slightly more additions than deletions. \\
Report sources & JIRA (50\%), GitHub (43.8\%), Google Code (6\%) & Provenance of natural-language reports. \\
\bottomrule
\end{tabular}
}
\vspace{0.5em}
\end{table}

To extend our analysis beyond curated Java benchmarks, we leverage SWT-Bench~\cite{mundler2024swt}, a dataset of real-world Python issues paired with fail-to-pass tests and human-written patches. We merge the \texttt{dev} and \texttt{test} splits (2,519 instances) and apply a reproducible cleaning and structuring pipeline. We parse patches into additions, deletions, and full diffs, reconstruct 
executable test functions, and retain only instances containing all three artifacts (bug report, test, patch), resulting in 2,346 aligned triplets. Metadata such as repository name, patch size, and test type are aggregated for statistical profiling. Table~\ref{tab:swtbench_stats} reports summary statistics.
We opt for SWT-Bench rather than the original SWE-bench~\cite{jimenez2023swe} because our study requires complete report–test–patch triplets with validated fail-to-pass tests explicitly linked to each issue. While SWE-bench provides real-world GitHub issues paired with human-written patches, it does not natively curate the triggering test associated with each issue as a first-class, validated artifact; SWT-Bench extends the same underlying issue corpus specifically to supply this missing test dimension, making it the more direct match for our cross-artifact alignment analysis.

\begin{table}[ht]
\vspace{0.5em}
\centering
\caption{Statistics of the curated \textsc{SWT-Bench} corpus.}
\label{tab:swtbench_stats}
\scalebox{0.7}{
\begin{tabular}{@{}p{4.2cm} p{4.6cm} p{8.8cm}@{}}
\toprule
\textbf{Metric} & \textbf{Value} & \textbf{Interpretation} \\
\midrule
Total instances (RAW) & 2,519 & All merged from Hugging Face (\texttt{dev} + \texttt{test}). \\
Triplets (bug--test--patch) & 2,346 (93.1\%) & Fully aligned artifacts across all repositories. \\
Unique repositories & 18 & Cross-project coverage across major Python ecosystems. \\
Avg.\ bug report length & 119.6 words & Issue textual richness and contextual diversity. \\
Avg.\ test length & 33.7 lines & Behavioral scope of fail--pass validation tests. \\
Avg.\ patch size & 41.2 lines & Typical granularity of human-written fixes. \\
Add/remove ratio & 4.48 & Balance between additive and corrective changes. \\
Executable tests & 2,346 (93.1\%) & Parsed Python test functions with assertions. \\
Declarative tests & 3 (0.1\%) & YAML or DSL-based test specifications. \\
\bottomrule
\end{tabular}
}
\vspace{0.5em}
\end{table}

Combined, Defects4J and SWT-Bench yield \textbf{2,857} report--test--patch triplets, forming the empirical basis for all subsequent analyses (Sec.~\ref{results}).

\subsection{Metrics and Evaluation}
\label{sub:metrics}

All analyses combine three complementary perspectives: (i)~\textbf{LLM-based evaluation}, (ii)~\textbf{metric-based similarity}, and (iii)~\textbf{human evaluation} for calibration. Evaluations are performed at the triplet level (report--test--patch) on both Defects4J and SWT-Bench to ensure cross-dataset comparability. LLM-based scores follow the four-dimension protocol 
described in Sec.~\ref{sub:llm_analysis} (coverage, correctness, specificity, overall alignment), produced independently by GPT-4o and DeepSeek-Chat and averaged across models, with justifications retained for qualitative analysis.

Metric-based evaluation covers lexical, semantic, and syntactic similarity between artifacts: lexical overlap via ROUGE-L and Jaccard; semantic similarity via cosine similarity over SBERT, CodeBERT, and OpenAI (\texttt{text-embedding-3-small}) embeddings; and structural similarity via POS-based noun and verb overlap using spaCy. We further analyze relationships between alignment scores and artifact-level features (e.g., report length, reproduction steps, exception mentions, test assertions) using non-parametric correlations (Spearman's~$\rho$, Kendall's~$\tau$, $p{<}0.05$), enabling cross-validation between LLM judgments and deterministic metrics. Table~\ref{tab:evaluation_metrics} summarizes all metrics and statistical procedures used throughout the study.

\begin{table*}[ht]
\vspace{0.5em}
\centering
\caption{Summary of the metrics and evaluation protocol used across all analyses.}
\label{tab:evaluation_metrics}
\scalebox{0.7}{%
\begin{tabular}{l p{4.2cm} p{8.8cm}}
\toprule
\textbf{Category} & \textbf{Metric / Dimension} & \textbf{Interpretation} \\
\midrule
\textbf{LLM-based} &
Coverage, Correctness, Specificity, Alignment (1–5) &
Expert ratings from GPT-4o and DeepSeek-Chat assessing how well artifacts (reports, tests, patches) align semantically and behaviorally. \\
\textbf{Lexical} &
ROUGE-L, Jaccard &
Measures surface textual overlap between natural-language descriptions and code artifacts, reflecting literal term reuse. \\
\textbf{Semantic} &
SBERT, CodeBERT, OpenAI\footnote{\url{https://platform.openai.com/docs/models/text-embedding-3-small}} + cosine similarity &
Captures conceptual relatedness beyond word overlap, linking bug descriptions, tests, and patches at the embedding level. \\
\textbf{Syntactic} &
POS overlap (verbs, nouns) via \textit{spaCy} &
Evaluates linguistic and structural correspondence, measuring overlap in action verbs and entities across artifacts. \\
% \textbf{Representation} &
% RAW vs.\ STRUCTURED reports &
% Compares natural-language input with structured representations (\textit{Entity}, \textit{Scenario}, \textit{Behavior}) extracted through LLM prompting. \\
\textbf{Report representation} &
RAW vs.\ STRUCTURED &
Compares the RAW natural-language report with the STRUCTURED representation (organized into
\textit{Entity}, \textit{Scenario}, and \textit{Behavior} fields) extracted through LLM prompting. \\
\textbf{Patch views} &
Full / Additions / Deletions &
Three complementary patch perspectives capturing the implementation, removal, or refactoring aspects of fixes. \\
\textbf{Statistical} &
Spearman’s~$\rho$, Kendall’s~$\tau$ (two-sided, $p{<}0.05$) &
Non-parametric correlations between textual and semantic features of reports/tests and their alignment quality. \\
\bottomrule
\end{tabular}
}
\vspace{0.5em}
\end{table*}

\subsection{Manual Verification and Quality Control}
\label{sub:manual_verification}

To anchor automated analyses in developer judgment, we conduct human evaluation on a shared sample of 400 bug reports (200 Defects4J, 200 SWT-Bench), covering the complete \textsc{Desc2Fix} pipeline across more than 29,000 individual field annotations and alignment ratings. Agreement across all human evaluation tasks, both structured-field annotation and alignment scoring, is quantified using the same set of non-parametric statistics: Spearman's $\rho$, Kendall's $\tau$, mean absolute error (MAE), and, for alignment scores, root mean squared error (RMSE) and Wilcoxon signed-rank tests with effect sizes.

Two annotators produced reference annotations for the six structured fields (exception, API, input conditions, reproduction steps, expected behavior, actual behavior), serving as ground truth for evaluating LLM-based extraction (precision, recall, F1); the resulting inter-annotator agreement is reported in Section~\ref{subsec:threats-to-validity}. Annotators also qualitatively inspected representative report--test and report--patch triplets to validate LLM alignment scores, identifying recurrent scoring behaviors (e.g., optimistic ratings for minimal patches, sensitivity to sparse tests) that informed minor prompt clarifications while preserving evaluation independence.

On the same sample, annotators independently rated report--test and report--patch behavioral alignment along coverage, correctness, specificity, and overall alignment. These ratings constitute the human reference for the inter-human and Human--LLM agreement analyses reported in our results. Report--test judgments were collected under RAW and STRUCTURED representations; report--patch judgments additionally covered Full, Add, and Remove patch views.

\subsection{Implementation and Configuration}
\label{sub:implementation_congig}
We conducted all experiments on a single workstation equipped with an Intel Core i9-14900K CPU (32 threads, 6.0 GHz), 64 GB RAM, and an NVIDIA RTX 5000 Ada GPU (32 GB VRAM). The entire pipeline was implemented in Python 3.10 and fully automated to enable scalable, reproducible execution across both Defects4J and SWT-Bench triplets.
LLM-based evaluations were performed with GPT-4o and DeepSeek-Chat via their official APIs under consistent decoding settings (temperature=1.0, top\_p=1.0), following the prompting scheme described in Section~\ref{sub:prompting_techniques}.
Semantic similarities were computed with SBERT (sentence-transformers/paraphrase-mpnet-base-v2\footnote{\url{https://huggingface.co/sentence-transformers/paraphrase-mpnet-base-v2}}) and CodeBERT (microsoft/codebert-base\footnote{\url{https://huggingface.co/microsoft/codebert-base}}) through the sentence-transformers framework, and with OpenAI's text-embedding-3-small via the OpenAI API. Cosine similarity was used on all embedding spaces. Lexical metrics include ROUGE-L and Jaccard. Syntactic alignment features were obtained with spaCy\footnote{\url{https://spacy.io/}} (English model) for POS tagging and lemmatization of verbs and nouns in reports, tests, and patches.

\section{Results and analysis}
\label{results} 

\subsection{RQ1: Reliability and Consistency of Structured Signal Extraction}
\label{subsec:RQ1}

\noindent \textbf{[Experimental design]:}  
RQ1 evaluates whether LLM-based structured extraction is reliable enough to serve as an enabling layer for alignment-driven debugging and repair analyses. Since RQ2 and RQ3 quantify cross-artifact alignment using structured report representations, extraction quality directly determines the validity of downstream signals.
Using Defects4J (Java) and SWT-Bench (Python), we compare GPT-4o and DeepSeek-Chat as independent extractors. Each model generates a JSON summary with six fields (exception, api\_involved, input\_conditions, reproduction\_steps, expected, actual), capturing actionable anchors across entity, scenario, and behavioral dimensions.
We conduct three complementary analyses to assess extraction robustness. The first, structural coverage, measures completeness as the proportion of non-null fields to characterize schema population behavior. The second, extraction accuracy, evaluates agreement with 400 manually annotated reports (200 per dataset) using precision, recall, and F1 under both completeness-aware and content-only regimes. The third, cross-model consistency, measures structural agreement and semantic similarity (ROUGE-L, Jaccard, SBERT) to assess the stability of extracted signals across independent systems.
Together, these analyses determine whether structured extraction provides a stable and semantically faithful representation suitable for alignment-based reasoning in downstream test generation, fault localization, and patch validation tasks.

\noindent \textbf{[Results]:} 

\noindent\textbf{\underline{RQ1.1 – Structural Coverage and Descriptive Baselines}}.
We first assess whether LLMs can populate our six-field schema at scale, as structural completeness directly determines how much report information becomes available for downstream alignment analyses (RQ2--RQ3). Table~\ref{tab:rq1-llm-global} shows that both GPT-4o and DeepSeek-Chat achieve consistently high completeness across corpora, as measured by the \emph{Compl.} column (the mean percentage of non-null fields per instance), with DeepSeek-Chat exhibiting a systematically denser extraction policy (Defects4J: 90.18\% vs.\ 86.14\%; SWT-Bench: 88.41\% vs.\ 83.48\%). The expected and actual behavioral anchors of the bug report (\textit{Expected} and \textit{Actual}) are near-saturated for both models across datasets ($\approx$98--100\%), indicating that behavioral outcome summaries are reliably populated regardless of extractor choice. Model differences instead concentrate on \emph{scenario specification} fields (\textit{API}, \textit{Input}, \textit{Repro}) and on \textit{Exception}: DeepSeek-Chat consistently populates scenario anchors more frequently (e.g., \textit{Repro} in SWT-Bench: 99.70\% vs.\ 75.92\%), suggesting a tendency to normalize implicit narrative descriptions into explicit procedural structure, whereas GPT-4o captures \textit{Exception} more often (Defects4J: 48.92\% vs.\ 44.03\%; SWT-Bench: 39.41\% vs.\ 32.48\%), consistent with a strategy that favors prominent, explicit error cues over less visible scenario detail.

\begin{table*}[ht]
\vspace{0.5em}
\centering
% \caption{Global summary of structured signal extraction by LLM and dataset (full corpora).}
\caption{Presence rates (\%) per field by LLM and dataset (full corpora).}
\label{tab:rq1-llm-global}
\scalebox{0.7}{
\begin{tabular}{@{}llccccccc@{}}
\toprule
\textbf{Dataset} & \textbf{Source} & \textbf{Compl.} & \textbf{Exception} & \textbf{API} & \textbf{Input} & \textbf{Repro} & \textbf{Expected} & \textbf{Actual} \\
\midrule
\multirow{2}{*}{Defects4J}
& GPT-4o     & 86.14 & \textbf{48.92} & 92.37 & 91.19 & 85.91 & 99.41 & 99.02 \\
& DeepSeek-Chat   & \textbf{90.18} & 44.03 & \textbf{98.83} & \textbf{99.22} & \textbf{99.22} & \textbf{99.80} & \textbf{100.00} \\
\addlinespace
\multirow{2}{*}{SWT-Bench}
& GPT-4o     & 83.48 & \textbf{39.41} & 95.00 & 92.40 & 75.92 & 99.62 & 98.51 \\
& DeepSeek-Chat   & \textbf{88.41} & 32.48 & \textbf{99.40} & \textbf{99.40} & \textbf{99.70} & \textbf{99.79} & \textbf{99.66} \\
\bottomrule
\end{tabular}
}
\begin{tablenotes}
\tiny
\centering \item[1]$^*$ Values are presence rates (\%) for each field; \emph{Compl.} is the mean percentage of the six fields that are non-null per instance.
\end{tablenotes}
\vspace{0.5em}
\end{table*}

To contextualize these baselines against human behavior, we report the same presence statistics on matched 400-report subsets (Table~\ref{tab:rq1-llm-sample100}). On both datasets, LLMs substantially exceed human completeness (e.g., Defects4J: 90.83\% DeepSeek-Chat / 86.17\% GPT-4o vs.\ 62--64\% for humans; SWT-Bench: 88.50\% / 83.33\% vs.\ 59--63\%). Rather than reflecting a genuine information advantage, this gap is consistent with LLMs being systematically optimistic extractors that rarely abstain: manual annotations reflect \emph{principled abstention} when a field is not explicitly supported by the report text, while LLMs more frequently infer and standardize implicit information (notably for \textit{Input} and \textit{Repro}) rather than declining to populate a field, a tendency that echoes the broader optimism bias we document for LLM-based alignment judgments later in this study (Sections~\ref{subsec:RQ2}--\ref{subsec:RQ3}) and that is consistent with related evidence that LLM-based agents can overstate their extraction of report-level anchors when reproducing failures~\cite{meng2024empirical}. Structured extraction is therefore sufficiently dense to support downstream alignment measurements, but extractors differ in their fill-versus-abstain policies: DeepSeek-Chat is preferable when maximizing coverage of scenario anchors (useful to scaffold test intents and reproduction structure), whereas exception-centric workflows may benefit from cross-validating the \textit{Exception} field with GPT-4o.

\begin{table*}[t]
\centering
\caption{Structured signal extraction on a representative sample of 400 bug reports.}
\label{tab:rq1-llm-sample100}
\scalebox{0.7}{
\begin{tabular}{@{}llccccccc@{}}
\toprule
\textbf{Dataset} & \textbf{Source} & \textbf{Compl. (\%)} & \textbf{Exception} & \textbf{API} & \textbf{Input} & \textbf{Repro} & \textbf{Expected} & \textbf{Actual} \\
\midrule
\multirow{4}{*}{Defects4J}
& GPT-4o     & 86.17 & \textbf{52.00} & 93.00 & 90.00 & 85.00 & 99.00 & 98.00 \\
& DeepSeek-Chat   & \textbf{90.83} & 48.00 & \textbf{98.00} & \textbf{99.00} & \textbf{100.00} & \textbf{100.00} & \textbf{100.00} \\
& Human-1    & 63.50 & 29.00 & 88.00 & 65.00 & 35.00 & 75.00 & 89.00 \\
& Human-2    & 62.33 & 34.00 & 84.00 & 63.00 & 32.00 & 72.00 & 89.00 \\
\addlinespace
\multirow{4}{*}{SWT-Bench}
& GPT-4o     & 83.33 & \textbf{41.00} & 96.00 & 92.00 & 74.00 & 100.00 & 97.00 \\
& DeepSeek-Chat   & \textbf{88.50} & 31.00 & \textbf{100.00} & \textbf{100.00} & \textbf{100.00} & \textbf{100.00} & \textbf{100.00} \\
& Human-1    & 62.83 & 21.00 & 82.00 & 64.00 & 39.00 & 74.00 & 97.00 \\
& Human-2    & 59.33 & 30.00 & 77.00 & 54.00 & 28.00 & 73.00 & 95.00 \\
\bottomrule
\end{tabular}
}
\begin{tablenotes}
\tiny
\centering \item[1]$^*$ Values are presence rates (\%) for each field; \emph{Compl.} is the average percentage of non-null fields per instance.
\end{tablenotes}
\vspace{3mm}
\end{table*}

\vspace{0.5em}
\noindent\colorbox{gray!15}{\parbox{0.98\linewidth}{
\textbf{Finding 1:} LLMs extract dense structured bug-report signals at scale (83--90\% completeness), but differ mainly by \emph{fill policy}. DeepSeek-Chat more often reconstructs scenario anchors (\textit{API/Input/Repro}), while GPT-4o more often surfaces \textit{Exception} evidence; Expected/Actual is near ceiling for both (98--100\%). Humans abstain more under ambiguity, while LLMs rarely decline to populate a field, suggesting LLM ``extra completeness'' reflects optimism rather than guaranteed evidence. Use LLM extraction as a reliable input contract, but treat scenario fields as coverage-oriented and exception fields as evidence-sensitive, with abstention rules or cross-model validation when needed.
}}
\vspace{0.5em}

\noindent\textbf{\underline{RQ1.2 – Extraction Accuracy}}.
We next evaluate how faithfully extracted fields match human references (Table~\ref{tab:rq1-acc-filtered}), distinguishing between two complementary regimes to disentangle semantic correctness from schema-completion policy. \emph{Completeness-aware accuracy} (Table~\ref{tab:rq1-acc-filtered}a) treats empty-versus-non-empty mismatches as errors, evaluating end-to-end extraction behavior including field population decisions. Under this regime, DeepSeek-Chat consistently outperforms GPT-4o (Defects4J: F1 = 0.595 vs.\ 0.514; SWT-Bench: 0.537 vs.\ 0.443), mirroring the denser population strategy observed in RQ1.1 rather than reflecting semantic disagreement.
In contrast, \emph{content-only fidelity} (Table~\ref{tab:rq1-acc-filtered}b), which evaluates only fields where both model and human provide non-empty content, reveals near-ceiling semantic agreement for both systems (F1 $\geq$ 0.93 across datasets): DeepSeek-Chat leads on Defects4J (0.963 vs.\ 0.931), while GPT-4o shows a marginal advantage on SWT-Bench (0.981 vs.\ 0.970). The collapse of the performance gap under this regime shows that when models commit to extracting a field, their semantic representations closely match human annotations.

This contrast demonstrates that most discrepancies originate from fill-versus-abstain decisions rather than semantic distortion: extraction variability primarily reflects population policy, not 
content hallucination. This distinction is critical from an engineering standpoint, as alignment signals computed over populated fields are unlikely to be degraded by semantic extraction errors, and precision--coverage trade-offs can instead be explicitly controlled via abstention policies or similarity thresholds. Practically, DeepSeek-Chat offers stronger end-to-end completeness-aware performance, suiting coverage-oriented pipelines, while GPT-4o remains slightly favored on SWT-Bench for evidence-sensitive, conservative extraction.

\begin{table}[ht]
\vspace{0.5em}
\centering
\caption{Extraction accuracy against human references with semantic filtering.}
\label{tab:rq1-acc-filtered}

\begin{subtable}{0.49\linewidth}
\centering
\caption{Completeness-aware accuracy (all matched pairs)}
\label{tab:rq1-acc-filtered-a}

\scalebox{0.7}{
\begin{tabular}{@{}llccc@{}}
\toprule
\textbf{Dataset} & \textbf{Model} & \textbf{P} & \textbf{R} & \textbf{F1} \\
\midrule
\multirow{2}{*}{Defects4J}
& GPT-4o   & 0.500 & 0.542 & 0.514 \\
& DeepSeek-Chat & \textbf{0.583} & \textbf{0.618} & \textbf{0.595} \\
\addlinespace
\multirow{2}{*}{SWT-Bench}
& GPT-4o   & 0.438 & 0.454 & 0.443 \\
& DeepSeek-Chat & \textbf{0.531} & \textbf{0.549} & \textbf{0.537} \\
\bottomrule
\end{tabular}
}

\end{subtable}
\hfill
\begin{subtable}{0.49\linewidth}
\centering
\caption{Non-empty pairs only}
\label{tab:rq1-acc-filtered-b}

\scalebox{0.7}{
\begin{tabular}{@{}llccc@{}}
\toprule
\textbf{Dataset} & \textbf{Model} & \textbf{P} & \textbf{R} & \textbf{F1} \\
\midrule
\multirow{2}{*}{Defects4J}
& GPT-4o   & 0.931 & 0.931 & 0.931 \\
& DeepSeek-Chat & \textbf{0.963} & \textbf{0.963} & \textbf{0.963} \\
\addlinespace
\multirow{2}{*}{SWT-Bench}
& GPT-4o   & \textbf{0.981} & \textbf{0.981} & \textbf{0.981} \\
& DeepSeek-Chat & 0.970 & 0.970 & 0.970 \\
\bottomrule
\end{tabular}
}

\end{subtable}
\vspace{0.5em}
\end{table}

\vspace{0.5em}
\noindent\colorbox{gray!15}{\parbox{0.98\linewidth}{
\textbf{Finding 2:} Extraction disagreements stem from field population policy, not semantic error. Under completeness-aware evaluation, DeepSeek-Chat achieves higher F1 due to more aggressive schema population. However, restricting evaluation to non-empty fields yields near-ceiling semantic fidelity for both models (F1 $\geq$ 0.93), indicating minimal semantic drift when content is produced. Structured extraction is therefore semantically stable, and downstream alignment analyses are unlikely to be biased by content-level hallucinations; coverage-precision trade-offs can be tuned via explicit abstention control.
}}
\vspace{0.5em}

\noindent\textbf{\underline{RQ1.3 – Model Consistency}}.
Beyond accuracy against human annotations, we evaluate whether structured extraction is stable across independent LLM systems. High inter-model agreement would indicate that alignment signals derived from structured reports are not overly sensitive to extractor choice. As shown in Table~\ref{tab:rq1-3-global-summary}, GPT-4o and DeepSeek-Chat make identical fill-versus-abstain decisions for the vast majority of fields (Defects4J: 93.61\%; SWT-Bench: 92.16\%), confirming that both systems converge on similar schema population decisions for most reports. When both models populate a field, semantic similarity is consistently high: mean ROUGE-L is approximately 0.65 across datasets, Jaccard ranges from 0.46 to 0.49, and SBERT cosine similarity reaches 0.80--0.81, indicating substantial semantic overlap despite lexical variation. DeepSeek-Chat maintains higher overall completeness (Defects4J: 90.25\% vs.\ 86.14\%, +4.11~pp; SWT-Bench: 88.43\% vs.\ 83.67\%, +4.76~pp), reflecting a more aggressive population strategy rather than divergent interpretation.

\vspace{0.5em}
\begin{table*}[ht]
\centering
\caption{Cross-model consistency overview on full corpora.}
\label{tab:rq1-3-global-summary}
\scalebox{0.7}{
\begin{tabular}{@{}l r ccc rrr@{}}
\toprule
\multirow{2}{*}{\textbf{Dataset}} & \multicolumn{1}{c}{\textbf{Structural}} & \multicolumn{3}{c}{\textbf{Semantic similarity (mean)}} & \multicolumn{3}{c}{\textbf{Completeness (\%)}} \\
\cmidrule(lr){2-2} \cmidrule(lr){3-5} \cmidrule(l){6-8}
& \textbf{Agreement (\%)} & ROUGE-L & Jaccard & SBERT cosine & GPT-4o & DeepSeek-Chat & $\Delta$ (Deep $-$ GPT) \\
\midrule
Defects4J & 93.61 & 0.647 & 0.459 & 0.804 & 86.14 & 90.25 & 4.11 \\
SWT-Bench  & 92.16 & 0.653 & 0.494 & 0.811 & 83.67 & 88.43 & 4.76 \\
\bottomrule
\end{tabular}
}
\begin{tablenotes}
\tiny
\centering \item Structural agreement (same non-null/empty decision per field), semantic similarity (mean ROUGE-L, Jaccard, SBERT cosine across fields), and average completeness by model (percentage of non-null fields).
\end{tablenotes}
%\vspace{0.5em}
\end{table*}

\vspace{0.5em}
\noindent\colorbox{gray!15}{\parbox{0.98\linewidth}{
\textbf{Finding 3:} Structured extraction is stable across independent LLM systems \emph{overall}. GPT-4o and DeepSeek-Chat agree on 92--94\% of fill-versus-abstain decisions and produce highly similar semantic content when both populate a field (SBERT $\approx$ 0.80). DeepSeek-Chat's 4--5~pp completeness advantage reflects denser schema population rather than semantic divergence. Consequently, alignment signals derived from structured reports are robust to extractor choice, and tuning coverage policies yields larger gains than switching models.
}}
\vspace{0.5em}

%%%%%%%%%%%
\begin{figure*}[t]
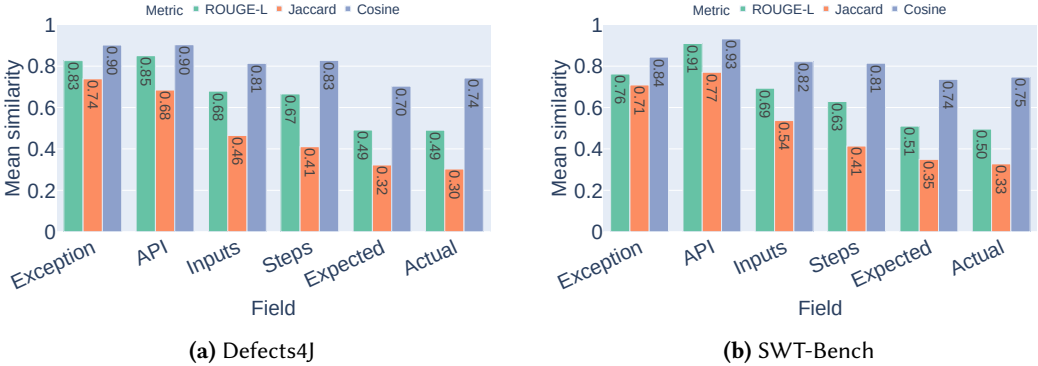

\centering

\begin{subfigure}{0.48\textwidth}
  \centering
  \includegraphics[width=\linewidth]{fig/RQ1_3_fig1_similarity_bars_Defects4J_.pdf}
  \caption{Defects4J}
  \label{fig:similarity-defects4j}
\end{subfigure}\hfill
\begin{subfigure}{0.48\textwidth}
  \centering
  \includegraphics[width=\linewidth]{fig/RQ1_3_fig1_similarity_bars_SWTBench_.pdf}
  \caption{SWT-Bench}
  \label{fig:similarity-swtbench}
\end{subfigure}

\caption{Semantic similarity between GPT-4o and DeepSeek-Chat by field and dataset.}
\label{fig:similarity_per_field}
\vspace{0.5em}
\end{figure*}

Breaking this down at the field level (Figure~\ref{fig:similarity_per_field}) reveals that consistency is not uniform. Entity-oriented fields (\textit{API}, \textit{Exception}) exhibit very high lexical and embedding similarity (ROUGE-L $\geq$ 0.83; cosine $\geq$ 0.90), suggesting near-canonical representations across models. Scenario fields (\textit{Input}, \textit{Repro}) show moderate lexical overlap but strong embedding similarity (cosine $\approx$ 0.81--0.83), consistent with paraphrased yet semantically equivalent procedural descriptions. Behavioral summaries (\textit{Expected}, \textit{Actual}) display the lowest lexical overlap (ROUGE-L $\approx$ 0.50) while maintaining moderate embedding similarity (cosine $\approx$ 0.70--0.75), reflecting natural summarization variability rather than semantic divergence. This gradient, from near-canonical entities to loosely paraphrased behaviors, confirms that inter-model differences primarily reflect field population policy rather than inconsistent semantic interpretation, and that extractor choice affects coverage density more than semantic content.

\vspace{0.5em}
\noindent\colorbox{gray!15}{\parbox{0.98\linewidth}{
\textbf{Finding 4:} Consistency is \emph{field-dependent}: entity fields (API, Exception) converge almost canonically across models, whereas scenario (Input, Repro) and behavioral (Expected, Actual) fields show progressively stronger paraphrasing, with lower ROUGE-L but moderate-to-high embedding similarity. Literal-overlap metrics therefore suffice for entity anchors, while embedding-based similarity is necessary to reliably compare scenario and behavioral signals.
}}
\vspace{0.5em}

RQ1 turns free-form bug reports into \emph{programmable artifacts}: dense, semantically faithful structured representations that automated tools can consume directly. We discuss the engineering implications of this finding, including concrete extraction-policy and ensemble-strategy guidance, in Section~\ref{subsec:eng-implications}.

\vspace{0.5em}
\noindent\highlight{Summary of \textbf{RQ1:} LLM-based extraction reliably converts bug reports into dense, machine-consumable behavioral anchors (83--90\% completeness), with DeepSeek-Chat and GPT-4o differing mainly in \emph{fill policy} rather than semantic accuracy (F1 $\geq$ 0.93 on populated fields; 92--94\% cross-model agreement). Structured anchors thus form a stable, model-robust input contract for RQ2--RQ3.
}
%\vspace{0.5em}

%%%%%%%%%%%%%%%%%%%%%%%%%%%%%%%%%%%%%%%%%%%%%%%%%%%%%%%%%%%%%%%%%%%%
\subsection{RQ2: Alignment Between Bug Descriptions and Triggering Tests}
\label{subsec:RQ2}

\noindent \textbf{[Experimental design]:} 
RQ2 asks whether a triggering test truly encodes the failure described in its bug report, and which report properties make that translation easier. We use Defects4J as the main corpus and SWT-Bench for cross-validation, comparing the RAW report against its STRUCTURED representation (extracted behavioral anchors: exception/API cues, input conditions, reproduction steps, expected vs.\ actual behavior).
We evaluate alignment with two complementary lenses: GPT-4o and DeepSeek-Chat score each report--test pair on a five-point scale along four axes (coverage, correctness, specificity, overall alignment), and we compute deterministic proxies, lexical overlap, embedding similarity, and verb/noun syntactic correspondence, that scale to ranking and filtering.
We quantify inter-human agreement as a reliability ceiling and Human--LLM agreement (Spearman's $\rho$, Kendall's $\tau$, absolute error) to expose optimism bias and ranking inconsistency, and we correlate anchor-derived report attributes (e.g., explicit expected/actual contrast, actionable steps, API/exception evidence, concise content) with both judge scores and metric proxies to connect alignment to report testability.

\noindent \textbf{[Results]:} 

\noindent\textbf{\underline{RQ2.1 – Bug Report–Test Semantic Alignment}}.
Table~\ref{tab:rq2_1_llm_aggregates} shows that alignment is \emph{non-trivial} on both datasets (means typically $>3/5$), but its magnitude is highly judge- and representation-dependent. Under RAW, DeepSeek-Chat is consistently more optimistic than GPT-4o, with the largest gaps on \textit{alignment} and \textit{correctness} (e.g., Defects4J alignment 3.81 vs.\ 3.25; correctness 3.98 vs.\ 3.26; SWT-Bench alignment 3.56 vs.\ 3.16; correctness 3.96 vs.\ 3.24), confirmed by Wilcoxon tests with medium-to-large paired effects (Defects4J: $r\approx+0.46$--$+0.48$; SWT-Bench: $r\approx+0.43$--$+0.58$; Table~\ref{tab:wilcoxon_model_effect_fullcorpus}). STRUCTURED inputs reduce DeepSeek-Chat scores and narrow or reverse this judge effect: DeepSeek-Chat drops markedly and becomes more variable (e.g., Defects4J alignment $3.03\pm1.62$ vs.\ $3.81\pm1.35$ in RAW), while GPT-4o remains comparatively stable and tends to yield higher \textit{specificity} (Defects4J: 3.51 vs.\ 3.28; SWT-Bench: 3.60 vs.\ 3.43); accordingly, Wilcoxon results reverse on Defects4J (GPT-4o $>$ DeepSeek-Chat on all axes, $|r|\approx0.10$--0.14) and become axis-dependent on SWT-Bench. Overall, LLM-based alignment is not an absolute quantity: RAW tends to inflate scores (especially for DeepSeek-Chat), whereas STRUCTURED yields more stable, specificity-oriented assessments better suited for ranking or filtering tests.

\begin{table*}[ht]
\vspace{0.5em}
\centering
%\caption{LLM-based alignment between bug reports and triggering tests (mean~$\pm$~std).}
\caption{LLM-based alignment between bug reports and triggering tests (mean~$\pm$~std). All scores are on a 1--5 scale.}
\label{tab:rq2_1_llm_aggregates}
\scalebox{0.6}{
\begin{tabular}{@{}lllcccc@{}}
\toprule
\textbf{Dataset} & \textbf{Scenario} & \textbf{Model} &
\textbf{Alignment} & \textbf{Coverage} & \textbf{Correctness} & \textbf{Specificity} \\
\midrule
\multirow{4}{*}{Defects4J}
& \multirow{2}{*}{RAW}        & GPT-4o   & 3.25~$\pm$~1.04 & 3.33~$\pm$~1.13 & 3.26~$\pm$~1.25 & 3.42~$\pm$~1.11 \\
&                             & DeepSeek-Chat & 3.81~$\pm$~1.35 & 3.64~$\pm$~1.49 & 3.98~$\pm$~1.43 & 3.92~$\pm$~1.28 \\
& \multirow{2}{*}{STRUCTURED} & GPT-4o   & 3.28~$\pm$~1.08 & 3.32~$\pm$~1.17 & 3.27~$\pm$~1.27 & 3.51~$\pm$~1.01 \\
&                             & DeepSeek-Chat & 3.03~$\pm$~1.62 & 3.12~$\pm$~1.68 & 2.98~$\pm$~1.74 & 3.28~$\pm$~1.58 \\
\addlinespace
\multirow{4}{*}{SWT-Bench}
& \multirow{2}{*}{RAW}        & GPT-4o   & 3.16~$\pm$~0.96 & 3.31~$\pm$~1.07 & 3.24~$\pm$~1.15 & 3.40~$\pm$~1.00 \\
&                             & DeepSeek-Chat & 3.56~$\pm$~1.09 & 3.49~$\pm$~1.19 & 3.96~$\pm$~1.19 & 3.77~$\pm$~1.10 \\
& \multirow{2}{*}{STRUCTURED} & GPT-4o   & 3.21~$\pm$~1.00 & 3.22~$\pm$~1.15 & 3.23~$\pm$~1.23 & 3.60~$\pm$~0.95 \\
&                             & DeepSeek-Chat & 3.16~$\pm$~1.26 & 3.40~$\pm$~1.34 & 3.26~$\pm$~1.48 & 3.43~$\pm$~1.21 \\
\bottomrule
\end{tabular}
}
\vspace{0.5em}
\end{table*}

\vspace{0.5em}
\noindent\colorbox{gray!15}{\parbox{0.98\linewidth}{
\textbf{Finding 5:} Report--test alignment is non-trivial but judge- and representation-dependent. Triggering tests generally receive alignment scores $>3/5$, yet absolute levels shift with both the evaluating LLM and the report view. Under RAW, DeepSeek-Chat scores are systematically higher than GPT-4o (medium-to-large paired effects), indicating a more optimistic judging style. Under STRUCTURED, DeepSeek-Chat scores drop and disperse, and the model gap narrows or can reverse (Defects4J: GPT-4o $>$ DeepSeek-Chat across axes; SWT-Bench: axis-dependent). Thus, alignment should be treated as a controllable signal: RAW tends to inflate scores, whereas STRUCTURED yields more stable, specificity-oriented assessments that are better suited for ranking and filtering tests.
}}
\vspace{0.5em}

Inter-human agreement is consistently strong across datasets and representations (Table~\ref{tab:inter_human_agreement}), establishing a reliable reference for alignment evaluation. On Defects4J, RAW correlations exceed $\rho=0.87$ for alignment and coverage and 
approach perfect agreement for correctness and specificity ($\rho \geq 0.99$); agreement decreases somewhat under STRUCTURED inputs (e.g., alignment $\rho$ from 0.87 to 0.71) but remains robust across axes. SWT-Bench follows a similar pattern (RAW: $\rho \in [0.82, 0.98]$; STRUCTURED: $\rho \geq 0.76$), with error magnitudes uniformly small ($MAE \leq 0.21$), an order of magnitude below model-level deviations. Human raters thus demonstrate stable ranking and consistent scoring behavior across representations.

\vspace{0.5em}
\begin{table*}[ht]
\centering
\caption{Inter-human agreement between bug reports and triggering tests}
\label{tab:inter_human_agreement}

\begin{subtable}{0.49\textwidth}
\centering
\caption{Defects4J}
\scalebox{0.7}{
\begin{tabular}{@{}lrrrrrrrr@{}}
\toprule
& \multicolumn{4}{c}{\textbf{RAW}} 
& \multicolumn{4}{c}{\textbf{STRUCTURED}} \\
\cmidrule(l){2-5} \cmidrule(l){6-9}
\textbf{Axis}
& $\rho$ & $\tau$ & MAE & RMSE
& $\rho$ & $\tau$ & MAE & RMSE \\
\midrule
Alignment   & 0.87 & 0.85 & 0.14 & 0.37 & 0.71 & 0.69 & 0.21 & 0.46 \\
Coverage    & 0.91 & 0.88 & 0.11 & 0.33 & 0.88 & 0.85 & 0.13 & 0.36 \\
Correctness & 1.00 & 1.00 & 0.00 & 0.00 & 0.99 & 0.99 & 0.02 & 0.14 \\
Specificity & 1.00 & 1.00 & 0.02 & 0.14 & 0.97 & 0.95 & 0.05 & 0.22 \\
\bottomrule
\end{tabular}
}
\end{subtable}
\hfill
\begin{subtable}{0.49\textwidth}
\centering
\caption{SWT-Bench}
\scalebox{0.7}{
\begin{tabular}{@{}lrrrrrrrr@{}}
\toprule
& \multicolumn{4}{c}{\textbf{RAW}} 
& \multicolumn{4}{c}{\textbf{STRUCTURED}} \\
\cmidrule(l){2-5} \cmidrule(l){6-9}
\textbf{Axis}
& $\rho$ & $\tau$ & MAE & RMSE
& $\rho$ & $\tau$ & MAE & RMSE \\
\midrule
Alignment   & 0.82 & 0.79 & 0.18 & 0.41 & 0.76 & 0.73 & 0.20 & 0.44 \\
Coverage    & 0.90 & 0.87 & 0.12 & 0.35 & 0.85 & 0.82 & 0.15 & 0.38 \\
Correctness & 0.98 & 0.96 & 0.03 & 0.17 & 0.95 & 0.92 & 0.06 & 0.25 \\
Specificity & 0.96 & 0.93 & 0.06 & 0.24 & 0.92 & 0.89 & 0.09 & 0.29 \\
\bottomrule
\end{tabular}
}
\end{subtable}

\vspace{2mm}
{\footnotesize $\rho$: Spearman correlation; $\tau$: Kendall correlation.}
\vspace{0.5em}
\end{table*}

Against this stable human baseline, Human--LLM agreement remains limited (Table~\ref{tab:rq2_human_llm_calibration_compact}). Rank correlations are weak across most settings: on Defects4J/RAW, $\bar{\rho}$ is near zero for GPT-4o (0.03) and negative for DeepSeek-Chat ($-0.13$), and while STRUCTURED modestly improves GPT-4o on Defects4J ($\bar{\rho}=0.25$) and SWT-Bench shows slightly higher correlations (up to 0.23), agreement remains far below 
inter-human levels throughout. Bias analysis reveals systematic score inflation: both models consistently over-score alignment relative to humans, with larger positive bias for DeepSeek-Chat (e.g., $\approx 1.99$ on Defects4J/RAW vs.\ 1.53 for GPT-4o), and model--human MAE (1.35--2.38) exceeds inter-human MAE ($\leq 0.21$) by an order of magnitude. Wilcoxon effect sizes are moderate to large ($\bar{r} \approx 0.58$--$0.77$), confirming substantial distributional shifts. In summary, while humans provide a stable ranking of report--test alignment, LLM judges exhibit optimistic bias and weak rank consistency; STRUCTURED inputs partially reduce variance for GPT-4o but do not eliminate calibration gaps.

%%%%%%%%%%%%%%%%%%%%%%%%%%%%%%%%%%%%%%%%%%%%%%%%%%%%%
\begin{table}[ht]
\vspace{0.5em}
\centering
\caption{Human--LLM agreement summary between bug reports and triggering tests.}
\label{tab:rq2_human_llm_calibration_compact}
\scalebox{0.7}{
\begin{tabular}{@{}lllrrrrrrrr@{}}
\toprule
\textbf{Dataset} & \textbf{Scenario} & \multicolumn{4}{c}{\textbf{GPT-4o}} & \multicolumn{4}{c}{\textbf{DeepSeek-Chat}} \\
\cmidrule(l){3-6} \cmidrule(l){7-10}
& & $\bar{\rho}$ & Bias & MAE & $\bar{r}$ & $\bar{\rho}$ & Bias & MAE & $\bar{r}$ \\
\midrule
Defects4J &
    RAW
 & 0.03 & 1.53 & 1.75 & 0.73 & -0.13 & 1.99 & 2.38 & 0.73 \\
 & 
    STRUCTURED
 & 0.25 & 1.59 & 1.71 & 0.77 & 0.02 & 1.45 & 1.90 & 0.58 \\
\addlinespace
SWT-Bench &
    RAW
 & 0.22 & 1.00 & 1.35 & 0.58 & 0.23 & 1.44 & 1.67 & 0.74 \\
 & 
    STRUCTURED
 & 0.09 & 1.51 & 1.67 & 0.75 & 0.08 & 1.63 & 1.88 & 0.73 \\
\addlinespace
\bottomrule
\end{tabular}
}
\begin{tablenotes}
\tiny
\centering \item Weighted averages computed using $n_{pairs}$ per axis. Spearman averages exclude undefined cases.
\end{tablenotes}
%\vspace{3mm}
\end{table}

%%%%%%%%%%%%%%%%%%%%%%%%%%%%%%%%%%%%%%%%%%%ùù

\vspace{0.5em}
\noindent\colorbox{gray!15}{\parbox{0.98\linewidth}{
\textbf{Finding 6:} Human test-adequacy ratings are stable, but LLM judges are optimistic and weakly rank-aligned. Inter-human agreement is high ($\rho > 0.8$, $MAE \leq 0.21$), whereas Human--LLM rank agreement is modest ($\bar{\rho}$ mostly in $[-0.13,\,0.25]$) with large positive bias and MAE up to 2.38. DeepSeek-Chat is more optimistic (especially in RAW), while GPT-4o improves slightly under STRUCTURED. Therefore, LLM judges should not be used out-of-the-box as decision signals; their scores require bias-aware calibration or should be complemented with deterministic proxies for ranking/filtering.
}}
\vspace{0.5em}

Deterministic metrics provide a consistent and reproducible view of report--test relatedness across datasets and representations (Tables~\ref{tab:rq2_1_metric_lexical_embedding} 
and~\ref{tab:rq2_1_metric_structured_signals}). Lexical overlap is uniformly minimal (ROUGE-L: 0.03--0.07; Jaccard: 0.03--0.06), confirming that triggering tests rarely reuse report phrasing 
verbatim and that surface-level matching is a poor indicator of alignment, whereas embedding-based similarity occupies a stable mid-range (SBERT: 0.47--0.54; OpenAI: 0.49--0.59), suggesting that tests capture report intent via semantic paraphrases rather than lexical copying. Across both corpora, RAW inputs yield higher embedding similarities than STRUCTURED, consistent with information loss when reports are condensed into extracted anchors. CodeBERT similarities are near-ceiling ($\approx$0.92--0.95), useful as a coarse ``recall'' sanity check but limited in discriminative power for ranking or filtering. Syntactic overlap further indicates that tests preferentially reuse \emph{entities} over \emph{actions}: POS overlaps are modest overall (verbs: 0.03--0.09; nouns: 0.06--0.22), with noun overlap consistently exceeding verb overlap and higher 
overall on Defects4J than SWT-Bench, suggesting stronger explicit entity reuse in Java-centric reports/tests. Finally, on structured dimensions (STRUCTURED-only), DeepSeek-Chat yields higher similarity than GPT-4o across entity, scenario, and behavior components, most notably on SWT-Bench scenario (0.468 vs.\ 0.392), indicating that structured fields can amplify semantic matching in narrative-style corpora.

\begin{table*}[ht]
\centering
\caption{Lexical and embedding similarities between bug reports and triggering tests.}
\label{tab:rq2_1_metric_lexical_embedding}

% ===================== (a) RAW =====================
\begin{subtable}{\textwidth}
\centering
\caption{RAW Inputs (Summary + Description)}
\label{tab:rq2_1_raw}

\vspace{1mm}

\scalebox{0.6}{
\begin{tabular}{llccccc}
\toprule
\textbf{Dataset} & \textbf{Model} & ROUGE-L & Jaccard & SBERT & CodeBERT & OpenAI \\
\midrule
\multirow{2}{*}{Defects4J}
& GPT-4o   & 0.070~$\pm$~0.087 & 0.062~$\pm$~0.071 & 0.538~$\pm$~0.130 & 0.948~$\pm$~0.031 & 0.591~$\pm$~0.096 \\
& DeepSeek-Chat & 0.069~$\pm$~0.085 & 0.062~$\pm$~0.070 & 0.537~$\pm$~0.130 & 0.949~$\pm$~0.031 & 0.589~$\pm$~0.096 \\
\addlinespace
\multirow{2}{*}{SWT-Bench}
& GPT-4o   & 0.037~$\pm$~0.037 & 0.033~$\pm$~0.032 & 0.523~$\pm$~0.120 & 0.925~$\pm$~0.042 & 0.524~$\pm$~0.105 \\
& DeepSeek-Chat & 0.037~$\pm$~0.039 & 0.033~$\pm$~0.032 & 0.524~$\pm$~0.119 & 0.925~$\pm$~0.041 & 0.525~$\pm$~0.105 \\
\bottomrule
\end{tabular}
}

\end{subtable}

\vspace{4mm}

% ===================== (b) STRUCTURED =====================
\begin{subtable}{\textwidth}
\centering
\caption{STRUCTURED Inputs (LLM-extracted signals)}
\label{tab:rq2_1_structured}

\vspace{1mm}

\scalebox{0.6}{
\begin{tabular}{llccccc}
\toprule
\textbf{Dataset} & \textbf{Model} & ROUGE-L & Jaccard & SBERT & CodeBERT & OpenAI \\
\midrule
\multirow{2}{*}{Defects4J}
& GPT-4o   & 0.032~$\pm$~0.033 & 0.032~$\pm$~0.032 & 0.473~$\pm$~0.120 & 0.934~$\pm$~0.025 & 0.555~$\pm$~0.099 \\
& DeepSeek-Chat & 0.033~$\pm$~0.031 & 0.035~$\pm$~0.029 & 0.503~$\pm$~0.114 & 0.944~$\pm$~0.021 & 0.577~$\pm$~0.092 \\
\addlinespace
\multirow{2}{*}{SWT-Bench}
& GPT-4o   & 0.036~$\pm$~0.038 & 0.032~$\pm$~0.032 & 0.469~$\pm$~0.127 & 0.919~$\pm$~0.040 & 0.495~$\pm$~0.112 \\
& DeepSeek-Chat & 0.038~$\pm$~0.037 & 0.034~$\pm$~0.032 & 0.493~$\pm$~0.126 & 0.926~$\pm$~0.045 & 0.518~$\pm$~0.112 \\
\bottomrule
\end{tabular}
}

\end{subtable}

\begin{tablenotes}
\tiny
\centering
\item Split by input scenario (RAW vs.\ STRUCTURED). All values are mean~$\pm$~std.
\end{tablenotes}

\end{table*}

\begin{table*}[ht]
\centering
\caption{POS overlaps and structured-dimension similarities between bug reports and triggering tests.}
\label{tab:rq2_1_metric_structured_signals}

\scalebox{0.6}{
\begin{tabular}{@{}llcc|cc|ccc@{}}
\toprule
\multirow{2}{*}{\textbf{Dataset}} & \multirow{2}{*}{\textbf{Model}} &
\multicolumn{2}{c|}{\textbf{POS (RAW)}} & \multicolumn{2}{c|}{\textbf{POS (STRUCTURED)}} & \multicolumn{3}{c}{\textbf{Structured Similarities}} \\
\cmidrule(lr){3-4} \cmidrule(lr){5-6} \cmidrule(l){7-9}
& & Verb ov. & Noun ov. & Verb ov. & Noun ov. & Entity & Scenario & Behavior \\
\midrule
\multirow{2}{*}{Defects4J}
& GPT-4o   & 0.055~$\pm$~0.112 & 0.124~$\pm$~0.164 & 0.045~$\pm$~0.144 & 0.127~$\pm$~0.194 & 0.386~$\pm$~0.146 & 0.370~$\pm$~0.146 & 0.334~$\pm$~0.114 \\
& DeepSeek-Chat & 0.092~$\pm$~0.127 & 0.217~$\pm$~0.172 & 0.077~$\pm$~0.159 & 0.213~$\pm$~0.157 & 0.418~$\pm$~0.123 & 0.432~$\pm$~0.116 & 0.361~$\pm$~0.118 \\
\addlinespace
\multirow{2}{*}{SWT-Bench}
& GPT-4o   & 0.026~$\pm$~0.078 & 0.057~$\pm$~0.100 & 0.035~$\pm$~0.110 & 0.084~$\pm$~0.143 & 0.383~$\pm$~0.150 & 0.392~$\pm$~0.163 & 0.394~$\pm$~0.126 \\
& DeepSeek-Chat & 0.027~$\pm$~0.080 & 0.057~$\pm$~0.102 & 0.030~$\pm$~0.104 & 0.076~$\pm$~0.130 & 0.425~$\pm$~0.129 & 0.468~$\pm$~0.127 & 0.424~$\pm$~0.125 \\
\bottomrule
\end{tabular}
}
\begin{tablenotes}
\tiny
\centering \item POS overlaps are computed on both RAW and STRUCTURED; structured similarities are only available for STRUCTURED inputs.
\end{tablenotes}
\vspace{2mm}
\end{table*}

\vspace{0.5em}
\noindent\colorbox{gray!15}{\parbox{0.98\linewidth}{
\textbf{Finding 7:} Triggering tests align mainly \emph{semantically}, not lexically. Report--test lexical overlap is near zero (ROUGE-L/Jaccard $\leq 0.07$), while embedding similarity is moderate and consistent (SBERT/OpenAI $\approx 0.47$--$0.59$), indicating paraphrased intent rather than textual copying (Table~\ref{tab:rq2_1_metric_lexical_embedding}). POS overlap is low and noun-dominated (nouns $>$ verbs), suggesting tests mirror \emph{entities} (APIs/identifiers) more than \emph{actions} (Table~\ref{tab:rq2_1_metric_structured_signals}). Consequently, ranking/filtering should favor semantic embeddings and entity anchors over lexical matching; improving testability requires making steps and expected/actual behavior explicit, not increasing verbosity.
}}
\vspace{0.5em}

OpenAI embedding cosine emerges as the most consistent deterministic predictor of LLM-based alignment judgments (Table~\ref{tab:rq2_1_llm_vs_metrics_top}). Across 15 out of 16 
dataset--scenario--model combinations, OpenAI cosine achieves the highest absolute Spearman correlation with LLM scores, with moderate but stable and statistically significant correlations (GPT-4o: $\rho \approx 0.32$--$0.45$; DeepSeek-Chat: $0.25$--$0.42$). Associations are generally stronger under STRUCTURED inputs, particularly for \textit{alignment} and \textit{coverage}, indicating that embedding similarity over distilled anchors captures the dimensions most emphasized by LLM judges. The only exception occurs for DeepSeek-Chat on Defects4J--STRUCTURED correctness, where \texttt{jaccard\_struct} slightly exceeds cosine, suggesting that token-level overlap may retain marginal utility when reasoning over condensed summaries. Overall, OpenAI embedding cosine provides a stable and representation-robust approximation of LLM alignment scores across datasets: unlike LLM judges, it is deterministic, inexpensive to compute, and free of optimistic bias drift.

\vspace{0.5em}
\begin{table*}[ht]
\centering
\caption{Best deterministic predictor per LLM axis (maximum $|\rho|$) by dataset, scenario, and model.}
\label{tab:rq2_1_llm_vs_metrics_top}
\scalebox{0.6}{
\begin{tabular}{@{}lllcccc@{}}
\toprule
\textbf{Dataset} & \textbf{Scenario} & \textbf{Model} &
\textbf{Alignment (best)} & \textbf{Coverage (best)} & \textbf{Correctness (best)} & \textbf{Specificity (best)} \\
\midrule
\multirow{4}{*}{Defects4J}
& \multirow{2}{*}{RAW}        & GPT-4o   & \texttt{openai\_raw} (0.382) & \texttt{openai\_raw} (0.387) & \texttt{openai\_raw} (0.375) & \texttt{openai\_raw} (0.394) \\
&                              & DeepSeek-Chat & \texttt{openai\_raw} (0.305) & \texttt{openai\_raw} (0.329) & \texttt{openai\_raw} (0.264) & \texttt{openai\_raw} (0.291) \\
& \multirow{2}{*}{STRUCTURED} & GPT-4o   & \texttt{openai\_struct} (0.452) & \texttt{openai\_struct} (0.425) & \texttt{openai\_struct} (0.353) & \texttt{openai\_struct} (0.421) \\
&                              & DeepSeek-Chat & \texttt{openai\_struct} (0.249) & \texttt{openai\_struct} (0.249) & \texttt{jaccard\_struct} (0.245) & \texttt{openai\_struct} (0.269) \\
\addlinespace
\multirow{4}{*}{SWT-Bench}
& \multirow{2}{*}{RAW}        & GPT-4o   & \texttt{openai\_raw} (0.388) & \texttt{openai\_raw} (0.394) & \texttt{openai\_raw} (0.326) & \texttt{openai\_raw} (0.348) \\
&                              & DeepSeek-Chat & \texttt{openai\_raw} (0.418) & \texttt{openai\_raw} (0.410) & \texttt{openai\_raw} (0.375) & \texttt{openai\_raw} (0.350) \\
& \multirow{2}{*}{STRUCTURED} & GPT-4o   & \texttt{openai\_struct} (0.393) & \texttt{openai\_struct} (0.408) & \texttt{openai\_struct} (0.318) & \texttt{openai\_struct} (0.338) \\
&                              & DeepSeek-Chat & \texttt{openai\_struct} (0.421) & \texttt{openai\_struct} (0.419) & \texttt{openai\_struct} (0.356) & \texttt{openai\_struct} (0.385) \\
\bottomrule
\end{tabular}
}

\vspace{2mm}
\end{table*}

\vspace{0.5em}
\noindent\colorbox{gray!15}{\parbox{0.98\linewidth}{
\textbf{Finding 8:} OpenAI embedding cosine is a stable and practical proxy for LLM alignment judgments. In 15/16 settings, OpenAI cosine is the strongest deterministic predictor of LLM scores (Table~\ref{tab:rq2_1_llm_vs_metrics_top}), with moderate but consistent correlations (up to $\rho \approx 0.45$). Associations are stronger under STRUCTURED inputs and for alignment/coverage axes. Embedding cosine can therefore replace LLM judges for ranking or filtering triggering tests, providing a reproducible, low-cost alignment signal without the optimistic bias observed in LLM evaluations.
}}
\vspace{0.5em}

\noindent\textbf{\underline{RQ2.2 – Bug Report Attributes and Test Quality}}.
We next examine which report-level attributes make failures easier to translate into executable triggering tests. Although correlations are generally small in magnitude (Tables~\ref{tab:rq2_llm_axis_top_attr} and~\ref{tab:rq2_1_top_attr_per_metric}), consistent patterns emerge across datasets.

\emph{Report length does not improve judged alignment.} On Defects4J, report length (\texttt{len\_report\_words}) exhibits systematic negative correlations with all LLM judgment axes in both RAW and STRUCTURED views ($\rho \approx -0.12$ to $-0.21$): longer descriptions increase narrative context but reduce judged alignment, suggesting a dilution effect where additional text does not translate into clearer, more testable behavioral cues. This pattern is corroborated by a recent study of LLM-based repair agents, which finds that longer bug reports are associated with lower resolution odds across SWE-bench Verified issues~\cite{khatib2026makes}, suggesting that narrative length is detrimental to actionability regardless of whether the outcome measure is judged alignment or downstream repair success. Notably, length-based features and \texttt{avg\_field\_length} (the mean word count across the six non-empty structured fields) correlate \emph{positively} with deterministic similarity metrics (e.g., ROUGE-L, SBERT, OpenAI cosine), indicating that longer reports increase textual or embedding overlap without improving perceived behavioral faithfulness.

\emph{Explicit behavioral anchors improve test alignment.} On SWT-Bench, LLM judgments are more strongly associated with the presence of explicit anchors than with length: the presence of APIs (\texttt{has\_api}) consistently improves alignment and coverage, and expected-outcome cues (\texttt{has\_expected}) increase specificity, particularly for GPT-4o. In contrast, deterministic similarity metrics remain largely driven by structural richness (e.g., number of entities, length) rather than by behavioral precision.

Taken together, LLM evaluators reward attributes that make report--test links behaviorally checkable (explicit APIs, steps, expected outcomes), whereas lexical or embedding similarity can increase mechanically with added text: tests align better with reports that expose \emph{operational anchors}, i.e., concrete, checkable elements such as named APIs, reproduction steps, and an explicit expected-versus-actual contrast, rather than narrative detail. This pattern may help explain a broader puzzle noted in the literature on LLM-based bug reproduction: recent evidence that LLM-based agents often fail to derive a working, failure-reproducing test directly from a bug report~\cite{nashid2025issue2test}, typically read as a limitation of the reproducing model. A report lacking explicit operational anchors may instead simply not carry enough behaviorally checkable information for any reproducer, human or automated, to act on.

\vspace{2mm}
\begin{table*}[ht]
\centering
\caption{Top report attribute per LLM judgement axis (Spearman $\rho$), grouped by dataset and model.}
\label{tab:rq2_llm_axis_top_attr}
\scalebox{0.6}{
\begin{tabular}{@{}lll llr llr@{}}
\toprule
\textbf{Dataset} & \textbf{Model} & \textbf{LLM axis} 
& \multicolumn{3}{c}{\textbf{RAW}} 
& \multicolumn{3}{c}{\textbf{STRUCTURED}} \\
\cmidrule(lr){4-6} \cmidrule(l){7-9}
& & & Best attr. & $\boldsymbol{\rho}$ & 
& Best attr. & $\boldsymbol{\rho}$ & \\
\midrule

\multirow{8}{*}{Defects4J} 
& \multirow{4}{*}{DeepSeek-Chat} 
    & Alignment   & \multirow{4}{*}{len\_report\_words} & $-0.197$ & & \multirow{4}{*}{len\_report\_words} & $-0.162$ & \\
&   & Correctness &   & $-0.147$ & & & $-0.151$ & \\
&   & Coverage    &   & $-0.178$ & & & $-0.145$ & \\
&   & Specificity &   & $-0.200$ & & & $-0.160$ & \\
\addlinespace

& \multirow{4}{*}{GPT-4o} 
    & Alignment   & \multirow{4}{*}{len\_report\_words} & $-0.208$ & & \multirow{4}{*}{len\_report\_words} & $-0.166$ & \\
&   & Correctness &   & $-0.160$ & & & $-0.168$ & \\
&   & Coverage    &   & $-0.160$ & & & $-0.123$ & \\
&   & Specificity &   & $-0.191$ & & & $-0.160$ & \\

\midrule

\multirow{8}{*}{SWT-Bench} 
& \multirow{4}{*}{DeepSeek-Chat} 
    & Alignment   & \multirow{4}{*}{avg\_field\_length} & $0.101$ & & has\_api           & $0.080$  & \\
&   & Correctness &   & $0.076$ & & avg\_field\_length & $0.079$  & \\
&   & Coverage    &   & $0.102$ & & has\_api           & $0.093$  & \\
&   & Specificity &   & $0.074$ & & avg\_field\_length & $0.041$  & \\
\addlinespace

& \multirow{4}{*}{GPT-4o} 
    & Alignment   & has\_exception     & $-0.051$ & & has\_api       & $0.082$  & \\
&   & Correctness & num\_entities      & $-0.084$ & & num\_entities  & $-0.081$ & \\
&   & Coverage    & has\_exception     & $-0.043$ & & has\_api       & $0.125$  & \\
&   & Specificity & has\_api           & $0.041$  & & has\_expected  & $0.074$  & \\

\bottomrule
\end{tabular}
}
\begin{tablenotes}
\tiny
\item[1] \textbf{len\_report\_words}: word count of RAW report (summary+description). \textbf{avg\_field\_length}: mean word count over non-empty structured fields. \textbf{has\_exception}: indicator that \textbf{exception} is present. \textbf{num\_entities}: count of distinct entities (exception/API names). \textbf{has\_api}: indicator that \textbf{api\_involved} is present. \textbf{has\_expected}: indicator that \textbf{expected} is present.
\end{tablenotes}

\vspace{2mm}
\end{table*}

\vspace{0.5em}
\noindent\colorbox{gray!15}{\parbox{0.98\linewidth}{
\textbf{Finding 9:} Testability depends more on explicit behavioral anchors than on report length. On Defects4J, longer reports (as measured by word count) increase lexical and embedding similarity but consistently reduce LLM alignment scores, indicating narrative dilution rather than added behavioral value. Across datasets, the presence of explicit APIs, reproduction steps, and expected outcomes more reliably predicts well-aligned triggering tests than increasing descriptive length. Bug reports may accordingly benefit from prioritizing concrete operational anchors (APIs, steps, expected behavior) over extended narrative context, though we do not test this design implication directly.
}}
\vspace{0.5em}

Findings 5--9 translate directly into design decisions for test-from-report pipelines. We operationalize these as a concrete alignment-guided test generation algorithm and discuss its engineering implications in Section~\ref{subsec:eng-implications}.

\vspace{0.5em}
\noindent\highlight{Summary of \textbf{RQ2:}
Report--test alignment is non-trivial but strongly judge- and representation-dependent: DeepSeek-Chat is more optimistic under RAW, while GPT-4o is more stable and specificity-oriented under
STRUCTURED. Human ratings form a reliable ceiling ($\rho > 0.8$), but both LLM judges show optimistic bias and weak rank agreement with humans, so their scores should not be used out-of-the-box. Alignment is predominantly semantic rather than lexical, with OpenAI embedding cosine the most consistent deterministic proxy for LLM judgments. Finally, explicit behavioral anchors, not report length, predict well-aligned tests; we build on this to design an alignment-guided test generation workflow in Section~\ref{subsec:eng-implications}.
}
%\vspace{0.5em}

%\vspace{0.5em}
\subsection{RQ3: Alignment Between Bug Descriptions and Corrective Patches}
\label{subsec:RQ3}

\noindent \textbf{[Experimental design]:} 
RQ3 examines how faithfully human-written patches translate the entities, scenarios, and behavioral contrasts described in bug reports into source-level repair actions. We analyze report--patch pairs by crossing two independent factors, report representation (RAW narratives vs.\ STRUCTURED anchors) and patch view (full diff, additions only, deletions only), yielding six conditions per dataset. Defects4J is our primary corpus and SWT-Bench provides cross-corpus validation.
Alignment is measured from two complementary perspectives. First, GPT-4o and DeepSeek-Chat act as judges and assign four 1--5 scores (overall alignment, coverage, correctness, specificity) with short justifications. Second, we compute deterministic proxies, lexical overlap (ROUGE-L, Jaccard), embedding similarity (SBERT, CodeBERT, OpenAI cosine), and POS-based verb/noun overlap, to separate reproducible similarity signals from the LLM-based judgments described above.
A human-rated subset provides a reliability ceiling and supports Human--LLM agreement analyses (rank agreement, bias, MAE, and Wilcoxon effect sizes) across these six conditions. Finally, we relate structured report attributes (e.g., length, reproduction steps, exception cues, expected/actual contrast) to patch alignment via Spearman's $\rho$ and Kendall's $\tau$, identifying which report semantics are associated with more behaviorally precise fixes. Together, this design characterizes patch directness as dependent on the report-representation/patch-view combination and quantifies how report semantics shape repair quality.

\noindent \textbf{[Results]:} 

\noindent\textbf{\underline{RQ3.1 – Directness of Patches to Behaviors}}.
As a reminder, a patch's \emph{full diff} is the complete change as committed, its \emph{additions} isolate only the lines it introduces, and its \emph{deletions} isolate only the lines it removes (Section~\ref{sub:datasets}). LLM judges generally rate human-written patches as well aligned with their bug reports, with the highest scores under RAW--full settings where mean alignment often exceeds 4/5 (Table~\ref{tab:rq3_llm_stats_by_rep_view}). Yet this perceived directness is not an intrinsic property of a patch--report pair: it varies systematically with both the \emph{patch view} and the \emph{report representation}. Across datasets and models, \emph{full diffs} provide the
strongest and most stable signal: add-only views preserve part of the implementation intent but drop corrective context, while remove-only views are the most fragile and representation-sensitive. Wilcoxon tests confirm that full typically outperforms add and remove, often with large effects ($|r|\!\geq\!0.5$; Table~\ref{tab:rq3_wilcoxon_view_effect}), and the penalty is strongest under the STRUCTURED representation, where deletions lose semantic anchoring (e.g., SWT-Bench full--remove: GPT-4o $r=-0.81^{***}$; DeepSeek-Chat $r=-0.70^{***}$).
Report representation amplifies these differences. RAW reports generally yield higher and less dispersed scores than STRUCTURED summaries, with the largest drops for DeepSeek-Chat: its Defects4J remove alignment falls from 4.66 (RAW) to 3.05 (STRUCTURED), consistent with higher sensitivity to abstraction and cue loss, whereas GPT-4o remains more robust (full specificity: 4.58 RAW vs.\ 4.52 STRUCTURED). Overall, patch--report directness depends on the report-representation/patch-view combination: full diffs stabilize evaluation by preserving context, while isolated add/remove views and the STRUCTURED representation reduce anchoring and expose model-specific sensitivities.

\vspace{0.6em}
%%%%%%%%%
\begin{table*}[ht]
\centering
\caption{LLM-based patch–report alignment scores (mean~$\pm$~sd, 1–5).}
\label{tab:rq3_llm_stats_by_rep_view}
\scalebox{0.7}{
\begin{tabular}{@{}lll lccc c@{}}
\toprule
\textbf{Dataset} & \textbf{Model} & \textbf{Repr.} & \textbf{View} &
\textbf{Alignment} & \textbf{Coverage} & \textbf{Correctness} & \textbf{Specificity} \\
\midrule

\multirow{12}{*}{Defects4J} 
& \multirow{6}{*}{GPT-4o} 
    & \multirow{3}{*}{RAW} 
        & add    & 3.16~$\pm$~1.37 & 3.38~$\pm$~1.26 & 3.07~$\pm$~1.39 & 3.50~$\pm$~1.38 \\
&&& full   & 3.97~$\pm$~1.24 & 4.46~$\pm$~0.91 & 4.05~$\pm$~1.21 & 4.58~$\pm$~0.79 \\
&&& remove & 3.82~$\pm$~1.22 & 4.10~$\pm$~1.14 & 3.91~$\pm$~1.30 & 4.24~$\pm$~1.12 \\
&& \multirow{3}{*}{STRUCTURED} 
        & add    & 2.58~$\pm$~1.05 & 2.46~$\pm$~1.16 & 2.59~$\pm$~1.27 & 3.65~$\pm$~1.13 \\
&&& full   & 3.69~$\pm$~1.05 & 3.67~$\pm$~1.15 & 3.71~$\pm$~1.29 & 4.52~$\pm$~0.75 \\
&&& remove & 2.95~$\pm$~1.08 & 2.81~$\pm$~1.18 & 2.99~$\pm$~1.30 & 3.87~$\pm$~1.10 \\

& \multirow{6}{*}{DeepSeek-Chat} 
    & \multirow{3}{*}{RAW} 
        & add    & 4.09~$\pm$~1.47 & 4.11~$\pm$~1.47 & 4.04~$\pm$~1.52 & 4.27~$\pm$~1.36 \\
&&& full   & 4.25~$\pm$~1.37 & 4.35~$\pm$~1.34 & 4.19~$\pm$~1.45 & 4.52~$\pm$~1.13 \\
&&& remove & 4.66~$\pm$~0.94 & 4.68~$\pm$~0.94 & 4.64~$\pm$~0.98 & 4.71~$\pm$~0.91 \\
&& \multirow{3}{*}{STRUCTURED} 
        & add    & 2.26~$\pm$~1.24 & 2.27~$\pm$~1.30 & 2.60~$\pm$~1.51 & 2.19~$\pm$~1.39 \\
&&& full   & 3.35~$\pm$~1.48 & 3.41~$\pm$~1.61 & 3.47~$\pm$~1.63 & 3.57~$\pm$~1.48 \\
&&& remove & 3.05~$\pm$~1.65 & 3.13~$\pm$~1.73 & 3.27~$\pm$~1.77 & 2.89~$\pm$~1.69 \\

\midrule

\multirow{12}{*}{SWT-Bench} 
& \multirow{6}{*}{GPT-4o} 
    & \multirow{3}{*}{RAW} 
        & add    & 4.00~$\pm$~1.13 & 4.33~$\pm$~0.90 & 4.21~$\pm$~1.02 & 4.43~$\pm$~0.91 \\
&&& full   & 4.17~$\pm$~1.02 & 4.54~$\pm$~0.79 & 4.46~$\pm$~0.88 & 4.64~$\pm$~0.74 \\
&&& remove & 2.89~$\pm$~1.33 & 3.17~$\pm$~1.34 & 2.79~$\pm$~1.42 & 3.49~$\pm$~1.37 \\
&& \multirow{3}{*}{STRUCTURED} 
        & add    & 3.73~$\pm$~1.08 & 3.71~$\pm$~1.18 & 3.88~$\pm$~1.17 & 4.29~$\pm$~0.98 \\
&&& full   & 4.10~$\pm$~0.99 & 4.08~$\pm$~1.08 & 4.23~$\pm$~1.03 & 4.52~$\pm$~0.82 \\
&&& remove & 2.45~$\pm$~1.04 & 2.25~$\pm$~1.14 & 2.39~$\pm$~1.25 & 3.35~$\pm$~1.20 \\

& \multirow{6}{*}{DeepSeek-Chat} 
    & \multirow{3}{*}{RAW} 
        & add    & 4.42~$\pm$~1.01 & 4.48~$\pm$~0.99 & 4.39~$\pm$~1.08 & 4.46~$\pm$~1.06 \\
&&& full   & 4.63~$\pm$~0.81 & 4.68~$\pm$~0.78 & 4.61~$\pm$~0.87 & 4.66~$\pm$~0.81 \\
&&& remove & 4.10~$\pm$~1.47 & 4.12~$\pm$~1.48 & 4.05~$\pm$~1.51 & 4.20~$\pm$~1.46 \\
&& \multirow{3}{*}{STRUCTURED} 
        & add    & 3.50~$\pm$~1.37 & 3.59~$\pm$~1.42 & 3.81~$\pm$~1.45 & 3.31~$\pm$~1.50 \\
&&& full   & 3.97~$\pm$~1.21 & 4.04~$\pm$~1.25 & 4.27~$\pm$~1.20 & 3.82~$\pm$~1.37 \\
&&& remove & 2.40~$\pm$~1.64 & 2.44~$\pm$~1.68 & 2.56~$\pm$~1.78 & 2.28~$\pm$~1.64 \\

\bottomrule
\end{tabular}
}
\vspace{2mm}
\end{table*}

\vspace{0.5em}
\noindent\colorbox{gray!15}{\parbox{0.98\linewidth}{
\textbf{Finding 10:} \emph{Behavioral directness} (how well a patch--report pair preserves the report's stated behavior) depends on the report-representation/patch-view combination, not on the pair alone. Full diffs yield the highest and most stable LLM alignment scores, whereas deletion-only views are systematically penalized, particularly under the STRUCTURED representation. RAW reports enhance robustness, especially for DeepSeek-Chat, while GPT-4o remains comparatively stable across formats.
}}
\vspace{0.5em}

Table~\ref{tab:inter_human_agreement_sample100} confirms that the human benchmark is highly reliable across patch views and representations. Inter-human agreement is near perfect for correctness and specificity (often $\rho \geq 0.99$ in RAW), consistently strong for coverage ($\rho > 0.80$), and moderate but stable for alignment, with MAE remaining small across settings (typically $\leq 0.44$), indicating that representation and patch view introduce difficulty but not incoherence among human raters, and establishing a robust upper bound for model comparison.
Against this stable baseline, Human--LLM rank agreement is generally modest (Table~\ref{tab:rq3_2_calibration_compact}). 

\emph{Rank agreement.} On Defects4J/RAW, average $\bar{\rho}$ remains near zero for most views (e.g., GPT-4o: 0.03; DeepSeek-Chat: $-0.08$ on full), with similar patterns on SWT-Bench/RAW. The clearest improvement appears under STRUCTURED+add on SWT-Bench ($\bar{\rho}=0.33$ for GPT-4o; 0.26 for DeepSeek-Chat), suggesting that structured behavioral cues combined with addition-focused diffs partially recover 
human-consistent ranking, though even in this best case, agreement remains far below inter-human levels.

\emph{Optimism bias.} LLMs consistently assign higher scores than humans: mean bias is positive across all datasets, representations, and views, frequently exceeding +2 points and peaking at +3.03 (DeepSeek-Chat, Defects4J/RAW-remove), an order of magnitude larger than inter-human MAE, indicating systematic inflation rather than random disagreement. Improvements in $\bar{\rho}$ do not imply closeness in absolute terms: MAE often remains above 2 in RAW settings and decreases only partially under STRUCTURED inputs. Wilcoxon effect sizes are consistently moderate to large ($\bar{r}$ typically 0.6--0.9 in RAW), confirming stable distributional 
divergence between human and LLM scores. Representation and patch view modulate agreement, but neither eliminates persistent optimism bias; LLM judgments should therefore be interpreted as 
condition-sensitive alignment signals rather than human-equivalent evaluations.

\begin{table*}[ht]
\vspace{0.5em}
\centering
\caption{Inter-human agreement between patches and bug reports}
\label{tab:inter_human_agreement_sample100}

\begin{subtable}{0.49\textwidth}
\centering
\caption{Defects4J}
\scalebox{0.7}{
\begin{tabular}{@{}llrrrrrr@{}}
\toprule
& & \multicolumn{3}{c}{\textbf{RAW}} 
& \multicolumn{3}{c}{\textbf{STRUCTURED}} \\
\cmidrule(l){3-5} \cmidrule(l){6-8}
\textbf{View} & \textbf{Axis}
& $\rho$ & $\tau$ & MAE
& $\rho$ & $\tau$ & MAE \\
\midrule

\multirow{4}{*}{Full}
& Align. & 0.75 & 0.71 & 0.34 & 0.39 & 0.37 & 0.44 \\
& Cov.   & 0.92 & 0.88 & 0.17 & 0.83 & 0.79 & 0.22 \\
& Corr.  & 0.99 & 0.97 & 0.09 & 1.00 & 1.00 & 0.00 \\
& Spec.  & 1.00 & 0.99 & 0.01 & 0.65 & 0.65 & 0.14 \\

\addlinespace

\multirow{4}{*}{Add}
& Align. & 0.58 & 0.56 & 0.41 & 0.48 & 0.47 & 0.41 \\
& Cov.   & 0.87 & 0.86 & 0.10 & 0.78 & 0.75 & 0.19 \\
& Corr.  & 1.00 & 1.00 & 0.01 & 1.00 & 1.00 & 0.00 \\
& Spec.  & 1.00 & 0.99 & 0.01 & 0.64 & 0.64 & 0.18 \\

\addlinespace

\multirow{4}{*}{Remove}
& Align. & 0.53 & 0.52 & 0.37 & 0.48 & 0.46 & 0.40 \\
& Cov.   & 0.84 & 0.82 & 0.14 & 0.78 & 0.73 & 0.28 \\
& Corr.  & 1.00 & 1.00 & 0.01 & 1.00 & 1.00 & 0.00 \\
& Spec.  & 0.99 & 0.98 & 0.01 & 0.51 & 0.51 & 0.14 \\

\bottomrule
\end{tabular}
}
\end{subtable}
\hfill
\begin{subtable}{0.49\textwidth}
\centering
\caption{SWT-Bench}
\scalebox{0.7}{
\begin{tabular}{@{}llrrrrrr@{}}
\toprule
& & \multicolumn{3}{c}{\textbf{RAW}} 
& \multicolumn{3}{c}{\textbf{STRUCTURED}} \\
\cmidrule(l){3-5} \cmidrule(l){6-8}
\textbf{View} & \textbf{Axis}
& $\rho$ & $\tau$ & MAE
& $\rho$ & $\tau$ & MAE \\
\midrule

\multirow{4}{*}{Full}
& Align. & 0.92 & 0.90 & 0.05 & 0.55 & 0.52 & 0.40 \\
& Cov.   & --   & --   & 0.00 & 0.89 & 0.83 & 0.26 \\
& Corr.  & 1.00 & 0.99 & 0.05 & 1.00 & 1.00 & 0.00 \\
& Spec.  & --   & --   & 0.00 & 0.85 & 0.84 & 0.10 \\

\addlinespace

\multirow{4}{*}{Add}
& Align. & 0.94 & 0.92 & 0.04 & 0.55 & 0.52 & 0.41 \\
& Cov.   & --   & --   & 0.00 & 0.89 & 0.84 & 0.25 \\
& Corr.  & 1.00 & 0.99 & 0.04 & 1.00 & 1.00 & 0.00 \\
& Spec.  & --   & --   & 0.00 & 0.85 & 0.84 & 0.10 \\

\addlinespace

\multirow{4}{*}{Remove}
& Align. & 0.98 & 0.98 & 0.01 & 0.55 & 0.53 & 0.43 \\
& Cov.   & --   & --   & 0.00 & 0.87 & 0.82 & 0.20 \\
& Corr.  & 1.00 & 1.00 & 0.01 & 1.00 & 1.00 & 0.00 \\
& Spec.  & --   & --   & 0.00 & 0.82 & 0.81 & 0.07 \\

\bottomrule
\end{tabular}
}
\end{subtable}

%\vspace{2mm}
{\footnotesize $\rho$: Spearman; $\tau$: Kendall. Dashes indicate undefined correlations due to constant inputs. Sample sizes vary by view (full: 200; add: 156/198; remove: 168/196)}
\vspace{0.5em}
\end{table*}
% {\footnotesize $\rho$: Spearman; $\tau$: Kendall. Dashes indicate undefined correlations due to constant inputs. Sample sizes vary by view (full: 100; add: 78/99; remove: 98/84).}

\begin{table}[ht]
\centering
\caption{Human--LLM patch–report agreement summary.}
\label{tab:rq3_2_calibration_compact}

\scalebox{0.7}{
\begin{tabular}{@{}lllrrrrrrrr@{}}
\toprule
\textbf{Dataset} & \textbf{Repr.} & \textbf{View} &
\multicolumn{4}{c}{\textbf{GPT-4o}} &
\multicolumn{4}{c}{\textbf{DeepSeek-Chat}} \\
\cmidrule(l){4-7} \cmidrule(l){8-11}
& & &
$\bar{\rho}$ & Bias & MAE & $\bar{r}$ &
$\bar{\rho}$ & Bias & MAE & $\bar{r}$ \\
\midrule

% ================= Defects4J =================
\multirow{6}{*}{Defects4J}
& \multirow{3}{*}{RAW}
& full   & 0.03 & 2.19 & 2.30 & 0.83 & -0.08 & 2.22 & 2.48 & 0.80 \\
& & add    & 0.16 & 1.36 & 1.64 & 0.67 & -0.07 & 2.28 & 2.48 & 0.80 \\
& & remove & 0.04 & 2.24 & 2.40 & 0.81 &  0.05 & 3.03 & 3.05 & 0.88 \\

\cmidrule(l){2-11}

& \multirow{3}{*}{STRUCT.}
& full   & 0.15 & 1.87 & 2.06 & 0.77 & 0.10 & 1.35 & 1.80 & 0.62 \\
& & add    & 0.11 & 1.09 & 1.34 & 0.62 & 0.12 & 0.66 & 1.06 & 0.38 \\
& & remove & 0.13 & 1.43 & 1.61 & 0.71 & 0.13 & 1.35 & 1.73 & 0.58 \\

\midrule

% ================= SWTBench =================
\multirow{6}{*}{SWT-Bench}
& \multirow{3}{*}{RAW}
& full   & -0.00 & 2.76 & 2.78 & 0.89 &  0.01 & 2.88 & 2.93 & 0.90 \\
& & add    & -0.04 & 2.49 & 2.56 & 0.86 & -0.01 & 2.70 & 2.78 & 0.88 \\
& & remove & -0.06 & 1.57 & 1.80 & 0.71 & -0.10 & 2.30 & 2.58 & 0.80 \\

\cmidrule(l){2-11}

& \multirow{3}{*}{STRUCT.}
& full   & 0.19 & 2.32 & 2.36 & 0.86 & 0.21 & 2.16 & 2.26 & 0.83 \\
& & add    & 0.33 & 2.04 & 2.12 & 0.84 & 0.26 & 1.67 & 1.86 & 0.76 \\
& & remove & 0.09 & 1.09 & 1.43 & 0.58 & 0.12 & 0.88 & 1.47 & 0.38 \\

\bottomrule
\end{tabular}
}

%\vspace{2mm}
{\footnotesize Weighted averages computed using $n_{pairs}$ per axis. Spearman averages exclude undefined cases.}
\vspace{0.5em}
\end{table}

\vspace{0.5em}
\noindent\colorbox{gray!15}{\parbox{0.98\linewidth}{
\textbf{Finding 11.} Human--LLM agreement depends on the report-representation/patch-view combination and is systematically optimistic. Rank correlations are modest and improve mainly under the STRUCTURED representation combined with the additions-only view on SWT-Bench, yet remain far below inter-human levels. LLMs consistently over-score alignment by 1--3 points, with large and stable distributional shifts. Agreement and bias must therefore be reported jointly, and LLM scores interpreted as condition-sensitive signals rather than human substitutes.
}}
\vspace{0.5em}

Across datasets and representations, full diffs consistently yield the strongest metric-based alignment, with add generally exceeding or matching remove (Table~\ref{tab:rq3_metric_stats_by_rep_view}). This pattern holds across lexical, embedding, and POS-based measures, confirming that contextual completeness provides the richest semantic 
signal. Transitioning from RAW to STRUCTURED introduces a systematic surface--semantics trade-off: embedding similarity declines (SBERT drops by 11--14\% on Defects4J full; OpenAI by 5--7\%), while predicate-oriented POS overlap increases substantially (verbs +28--54\%, nouns +16--21\%), and lexical dispersion contracts while embedding variance remains stable or slightly higher, indicating reduced surface redundancy but preserved semantic diversity. This 
RAW$\to$STRUCTURED drop is more pronounced on Defects4J than SWT-Bench, suggesting stronger reliance on lexical/contextual richness in Defects4J reports, whereas SWT-Bench maintains more stable embedding alignment under abstraction. CodeBERT, by contrast, remains consistently high ($\sim$0.92--0.96) and largely insensitive to representation or view, evidencing stable code-level correspondence independent of surface framing. Overall, metric-based evidence corroborates the framing effects observed with LLM judges: full diffs maximize contextual signal, whereas structured abstraction shifts alignment from surface similarity toward predicate--argument salience.

\vspace{0.5em}
\begin{table*}[ht]
\centering
\caption{Metric-based patch--report alignment scores (mean~$\pm$~sd, 0--1).}
\label{tab:rq3_metric_stats_by_rep_view}
\scalebox{0.7}{
\begin{tabular}{@{}ll l lccccccc@{}}
\toprule
\textbf{Dataset} & \textbf{Repr.} & \textbf{View} &
\textbf{ROUGE-L} & \textbf{Jaccard} & \textbf{SBERT} & \textbf{CodeBERT} & \textbf{OpenAI} & \textbf{Verbs} & \textbf{Nouns} \\
\midrule
\multirow{6}{*}{Defects4J}
  & \multirow{3}{*}{RAW}
      & add    & 0.02~$\pm$~0.03 & 0.02~$\pm$~0.02 & 0.37~$\pm$~0.13 & 0.91~$\pm$~0.04 & 0.39~$\pm$~0.12 & 0.02~$\pm$~0.05 & 0.03~$\pm$~0.04 \\
  &&  full   & 0.06~$\pm$~0.05 & 0.05~$\pm$~0.04 & 0.58~$\pm$~0.11 & 0.96~$\pm$~0.02 & 0.58~$\pm$~0.11 & 0.06~$\pm$~0.08 & 0.14~$\pm$~0.11 \\
  &&  remove & 0.03~$\pm$~0.04 & 0.02~$\pm$~0.03 & 0.41~$\pm$~0.13 & 0.92~$\pm$~0.03 & 0.42~$\pm$~0.12 & 0.03~$\pm$~0.06 & 0.03~$\pm$~0.05 \\
  & \multirow{3}{*}{STRUCTURED}
      & add    & 0.01~$\pm$~0.03 & 0.01~$\pm$~0.02 & 0.33~$\pm$~0.14 & 0.92~$\pm$~0.05 & 0.38~$\pm$~0.14 & 0.04~$\pm$~0.13 & 0.03~$\pm$~0.06 \\
  &&  full   & 0.04~$\pm$~0.03 & 0.04~$\pm$~0.03 & 0.50~$\pm$~0.11 & 0.95~$\pm$~0.04 & 0.55~$\pm$~0.13 & 0.10~$\pm$~0.16 & 0.17~$\pm$~0.14 \\
  &&  remove & 0.02~$\pm$~0.03 & 0.01~$\pm$~0.02 & 0.37~$\pm$~0.14 & 0.93~$\pm$~0.04 & 0.41~$\pm$~0.13 & 0.05~$\pm$~0.13 & 0.04~$\pm$~0.07 \\
\midrule
\multirow{6}{*}{SWT-Bench}
  & \multirow{3}{*}{RAW}
      & add    & 0.04~$\pm$~0.04 & 0.04~$\pm$~0.04 & 0.45~$\pm$~0.14 & 0.92~$\pm$~0.04 & 0.43~$\pm$~0.13 & 0.06~$\pm$~0.11 & 0.10~$\pm$~0.12 \\
  &&  full   & 0.04~$\pm$~0.04 & 0.04~$\pm$~0.04 & 0.47~$\pm$~0.13 & 0.93~$\pm$~0.04 & 0.44~$\pm$~0.12 & 0.06~$\pm$~0.11 & 0.13~$\pm$~0.13 \\
  &&  remove & 0.03~$\pm$~0.04 & 0.03~$\pm$~0.03 & 0.40~$\pm$~0.14 & 0.92~$\pm$~0.04 & 0.39~$\pm$~0.12 & 0.03~$\pm$~0.08 & 0.06~$\pm$~0.09 \\
  & \multirow{3}{*}{STRUCTURED}
      & add    & 0.04~$\pm$~0.04 & 0.04~$\pm$~0.04 & 0.42~$\pm$~0.15 & 0.92~$\pm$~0.05 & 0.42~$\pm$~0.13 & 0.07~$\pm$~0.15 & 0.13~$\pm$~0.16 \\
  &&  full   & 0.04~$\pm$~0.04 & 0.04~$\pm$~0.03 & 0.42~$\pm$~0.13 & 0.93~$\pm$~0.05 & 0.42~$\pm$~0.12 & 0.08~$\pm$~0.15 & 0.15~$\pm$~0.16 \\
  &&  remove & 0.03~$\pm$~0.04 & 0.02~$\pm$~0.03 & 0.37~$\pm$~0.15 & 0.93~$\pm$~0.05 & 0.37~$\pm$~0.13 & 0.03~$\pm$~0.10 & 0.07~$\pm$~0.11 \\
\bottomrule
\end{tabular}
}
\vspace{2mm}
\end{table*}

\vspace{0.5em}
\noindent\colorbox{gray!15}{\parbox{0.98\linewidth}{
\textbf{Finding 12:} Lexical, embedding, and syntactic similarity metrics confirm the same report-representation/patch-view effects observed with LLM judges. Full diffs yield the highest alignment across lexical and embedding measures. The STRUCTURED representation reduces surface and embedding similarity but increases predicate-oriented POS overlap, indicating a shift from surface fidelity to argument-level salience. This trade-off is stronger on Defects4J than SWT-Bench. CodeBERT remains near-ceiling and representation-invariant, providing a stable code-level anchor.
}}
\vspace{0.5em}

\noindent\textbf{\underline{RQ3.2 – Report Semantics and Patch Precision}}.
Associations between report attributes and LLM-judged patch quality display corpus-dependent patterns (Table~\ref{tab:rq3_2_llm_dir_summary}). On Defects4J, GPT-4o shows 
virtually no stable associations, whereas DeepSeek-Chat exhibits coherent but small negative correlations across all axes (median $\rho$ in the $-0.05$ to $-0.09$ range): increased procedural or lexical density (more steps, entities, longer fields) slightly reduces perceived alignment, coverage, and specificity. Effect sizes remain modest ($|\rho| \leq 0.08$) but directionally consistent. On SWT-Bench, both judges predominantly reward behavioral contrast 
instead: explicit expected/actual statements and concise, content-bearing fields positively associate with alignment, coverage, and specificity (median $\rho \approx 0.03$--$0.05$), and 
DeepSeek-Chat mirrors GPT-4o's positive trends while additionally penalizing excessive input enumeration (negative correlations for \texttt{num\_inputs}). Overall, clarity and contrast outweigh sheer descriptive volume: verbosity is weakly detrimental on Defects4J (DeepSeek-Chat), whereas explicit behavioral contrast improves precision on SWT-Bench, small but internally consistent, corpus-specific sensitivities.

%%%%%%%%%%%%%%%%%%%%%%%%%%%%%%%%%
\begin{table*}[ht]
\vspace{0.5em}
\centering
\caption{Direction and magnitude of LLM-based correlations between \emph{report attributes} and patch-quality scores.}
\label{tab:rq3_2_llm_dir_summary}

% ========== (a) Defects4J ==========
\textbf{(a) Defects4J}
\vspace{1mm}

\scalebox{0.6}{
\begin{tabular}{@{}llrrrcc@{}}
\toprule
\textbf{Score} & \textbf{Model} & \textbf{Sig +} & \textbf{Sig --} & \textbf{Median $\rho$ (sig)} & \textbf{Median $\tau$ (sig)} & \textbf{Max $|\rho|$ (sig)} \\
\midrule
\multirow{2}{*}{Alignment}
 & DeepSeek-Chat & 0 & 4 & –0.072 & –0.056 & 0.044 \\
 & GPT-4o   & 0 & 0 & –      & –      & –     \\
\multirow{2}{*}{Correctness}
 & DeepSeek-Chat & 0 & 4 & –0.085 & –0.068 & 0.052 \\
 & GPT-4o   & 0 & 0 & –      & –      & –     \\
\multirow{2}{*}{Coverage}
 & DeepSeek-Chat & 0 & 4 & –0.079 & –0.063 & 0.042 \\
 & GPT-4o   & 0 & 1 & –0.044 & –0.033 & 0.033 \\
\multirow{2}{*}{Specificity}
 & DeepSeek-Chat & 0 & 3 & –0.054 & –0.047 & 0.031 \\
 & GPT-4o   & 0 & 0 & –      & –      & –     \\
\bottomrule
\end{tabular}
}

%\vspace{4mm}

% ========== (b) SWTBench ==========
\textbf{(b) SWT-Bench}
\vspace{1mm}

\scalebox{0.6}{
\begin{tabular}{@{}llrrrcc@{}}
\toprule
\textbf{Score} & \textbf{Model} & \textbf{Sig +} & \textbf{Sig --} & \textbf{Median $\rho$ (sig)} & \textbf{Median $\tau$ (sig)} & \textbf{Max $|\rho|$ (sig)} \\
\midrule
\multirow{2}{*}{Alignment}
 & DeepSeek-Chat & 5 & 1 & 0.031 & 0.027 & 0.079 \\
 & GPT-4o   & 6 & 0 & 0.042 & 0.036 & 0.074 \\
\multirow{2}{*}{Correctness}
 & DeepSeek-Chat & 6 & 1 & 0.032 & 0.030 & 0.055 \\
 & GPT-4o   & 3 & 0 & 0.048 & 0.039 & 0.052 \\
\multirow{2}{*}{Coverage}
 & DeepSeek-Chat & 5 & 1 & 0.026 & 0.022 & 0.082 \\
 & GPT-4o   & 7 & 0 & 0.045 & 0.041 & 0.068 \\
\multirow{2}{*}{Specificity}
 & DeepSeek-Chat & 3 & 1 & 0.032 & 0.026 & 0.073 \\
 & GPT-4o   & 7 & 0 & 0.046 & 0.039 & 0.055 \\
\bottomrule
\end{tabular}
}
\begin{tablenotes}
\tiny
\item \textbf{Sig +} / \textbf{Sig --}: count of report attributes with significant positive / negative correlation (significance if Spearman $p<.05$ or Kendall $p<.05$).
\item \textbf{Median $\rho$ (sig)} / \textbf{Median $\tau$ (sig)}: median Spearman / Kendall among significant effects within the block.
\item \textbf{Max $|\rho|$}: largest absolute Spearman coefficient among significant effects in the block.
\end{tablenotes}

\vspace{0.5em}
\end{table*}

\vspace{0.5em}
\noindent\colorbox{gray!15}{\parbox{0.98\linewidth}{
\textbf{Finding 13:} Report--patch precision is corpus- and judge-dependent. On SWT-Bench, explicit behavioral contrast (expected/actual cues, concise salient steps) consistently improves LLM-judged alignment. On Defects4J, increased procedural or lexical density weakly reduces DeepSeek-Chat scores, while GPT-4o remains largely insensitive. Clarity and contrast matter more than volume, but their impact depends on corpus characteristics.
}}
\vspace{0.5em}

Correlations between report attributes and metric-based report--patch similarity also exhibit corpus-dependent structure (Table~\ref{tab:rq3_2_metric_dir_summary}). On Defects4J, associations 
are modest and mixed: code-aware embeddings (SBERT, CodeBERT) account for most positive effects, whereas surface metrics (ROUGE-L, Jaccard) show sparse or unstable associations, and POS-based overlaps frequently tilt negative, indicating that simple lexical or predicate cues can invert under this corpus' reporting style. On SWT-Bench, the pattern is substantially more uniform: lexical, embedding, and POS metrics predominantly show positive correlations, especially for 
verbs and code-aware embeddings, and although effect sizes remain small (median $|\rho| \approx 0.03$--0.08), their directional consistency suggests stable semantic reinforcement when reports are clearer and more content-bearing. Across both datasets, SBERT and CodeBERT provide the most consistent positive associations, while surface and POS metrics are corpus-sensitive, reinforcing the need to report multiple metric families per dataset rather than relying on a single similarity signal.

\begin{table*}[ht]
\vspace{0.5em}
\centering
\caption{Direction and magnitude of metric-based correlations between report attributes and patch alignment metrics.}
\label{tab:rq3_2_metric_dir_summary}

% ========== (a) Defects4J ==========
\textbf{(a) Defects4J}
\vspace{1mm}

\scalebox{0.6}{
\begin{tabular}{@{}llrrrcc@{}}
\toprule
\textbf{Metric} & \textbf{Model} & \textbf{Sig +} & \textbf{Sig --} & \textbf{Median $\rho$ (sig)} & \textbf{Median $\tau$ (sig)} & \textbf{Max $|\rho|$} \\
\midrule
\multirow{2}{*}{CodeBERT}
 & GPT-4o   & 7  & 0 &  0.066   & 0.054   & 0.125 \\
 & DeepSeek-Chat & 2  & 0 &  0.058   & 0.043   & 0.068 \\
\multirow{2}{*}{Jaccard}
 & GPT-4o   & 2  & 0 &  0.068   & 0.052   & 0.082 \\
 & DeepSeek-Chat & 1  & 1 &  0.018   & 0.012   & 0.089 \\
\multirow{2}{*}{Noun overlap}
 & GPT-4o   & 1  & 1 & –0.0065  & –0.0025 & 0.071 \\
 & DeepSeek-Chat & 0  & 2 & –0.0575  & –0.043  & 0.060 \\
\multirow{2}{*}{OpenAI (emb)}
 & GPT-4o   & 6  & 3 &  0.057   & 0.045   & 0.155 \\
 & DeepSeek-Chat & 2  & 3 & –0.054   & –0.041  & 0.091 \\
\multirow{2}{*}{ROUGE-L}
 & GPT-4o   & 2  & 0 &  0.065   & 0.051   & 0.072 \\
 & DeepSeek-Chat & 1  & 2 & –0.072   & –0.058  & 0.101 \\
\multirow{2}{*}{SBERT}
 & GPT-4o   & 10 & 0 &  0.075   & 0.055   & 0.227 \\
 & DeepSeek-Chat & 4  & 0 &  0.113   & 0.081   & 0.169 \\
\multirow{2}{*}{Verb overlap}
 & GPT-4o   & 1  & 2 & –0.054   & –0.043  & 0.080 \\
 & DeepSeek-Chat & 0  & 1 & –0.066   & –0.056  & 0.066 \\
\bottomrule
\end{tabular}
}

%\vspace{4mm}

% ========== (b) SWTBench ==========
\textbf{(b) SWT-Bench}
\vspace{1mm}

\scalebox{0.6}{
\begin{tabular}{@{}llrrrcc@{}}
\toprule
\textbf{Metric} & \textbf{Model} & \textbf{Sig +} & \textbf{Sig --} & \textbf{Median $\rho$ (sig)} & \textbf{Median $\tau$ (sig)} & \textbf{Max $|\rho|$} \\
\midrule
\multirow{2}{*}{CodeBERT}
 & GPT-4o   & 11 & 0 &  0.079   & 0.060   & 0.160 \\
 & DeepSeek-Chat & 2  & 0 &  0.059   & 0.048   & 0.060 \\
\multirow{2}{*}{Jaccard}
 & GPT-4o   & 9  & 0 &  0.064   & 0.053   & 0.131 \\
 & DeepSeek-Chat & 10 & 0 &  0.057   & 0.048   & 0.187 \\
\multirow{2}{*}{Noun overlap}
 & GPT-4o   & 6  & 5 &  0.030   & 0.026   & 0.112 \\
 & DeepSeek-Chat & 6  & 5 &  0.042   & 0.036   & 0.139 \\
\multirow{2}{*}{OpenAI (emb)}
 & GPT-4o   & 9  & 3 &  0.050   & 0.039   & 0.180 \\
 & DeepSeek-Chat & 9  & 2 &  0.074   & 0.057   & 0.225 \\
\multirow{2}{*}{ROUGE-L}
 & GPT-4o   & 14 & 0 &  0.043   & 0.035   & 0.123 \\
 & DeepSeek-Chat & 11 & 0 &  0.062   & 0.051   & 0.165 \\
\multirow{2}{*}{SBERT}
 & GPT-4o   & 11 & 0 &  0.068   & 0.056   & 0.146 \\
 & DeepSeek-Chat & 12 & 0 &  0.065   & 0.053   & 0.207 \\
\multirow{2}{*}{Verb overlap}
 & GPT-4o   & 5  & 3 &  0.040   & 0.037   & 0.123 \\
 & DeepSeek-Chat & 8  & 1 &  0.037   & 0.032   & 0.158 \\
\bottomrule
\end{tabular}
}

%\vspace{1mm}

\begin{tablenotes}
\tiny
\item \textbf{Sig +} / \textbf{Sig --}: count of report attributes with significant positive / negative correlation (significance if Spearman $p<.05$ or Kendall $p<.05$).
\item \textbf{Median $\rho$ (sig)} / \textbf{Median $\tau$ (sig)}: median Spearman / Kendall among significant effects within the block.
\item \textbf{Max $|\rho|$}: largest absolute Spearman coefficient among significant effects in the block.
\end{tablenotes}

\vspace{2mm}
\end{table*}

\vspace{0.5em}
\noindent\colorbox{gray!15}{\parbox{0.98\linewidth}{
\textbf{Finding 14:} Correlations between report attributes and metric-based alignment are small but structurally consistent. Code-aware embeddings (SBERT, CodeBERT) show stable positive associations between informative reports and patch similarity across datasets. Surface and POS metrics are corpus-sensitive, mixed or negative on Defects4J, but uniformly positive on SWT-Bench. Embeddings therefore provide reliable primary signals, while lexical/POS cues offer complementary, dataset-dependent evidence.
}}
\vspace{0.5em}

Findings 10--14 show that patch--report alignment is representation-sensitive and signal-dependent rather than intrinsic. We operationalize these into a concrete alignment-guided patch generation and validation algorithm, discussed alongside its engineering implications in Section~\ref{subsec:eng-implications}.

\vspace{1.5em}
\noindent\highlight{Summary of \textbf{RQ3:} Patch--report alignment depends strongly on the report-representation/patch-view combination: full diffs provide the richest and most stable signal, RAW representations inflate scores, and STRUCTURED representations sharpen behavioral focus at the cost of surface similarity. Human judgments remain stable across views, but LLM judges show systematic optimism (+1--2 points) and only modest rank agreement with humans, making them unsuitable as out-of-the-box decision signals. Embedding-based similarity (SBERT, CodeBERT, OpenAI) offers the most stable cross-dataset proxy, while explicit behavioral contrast, not report length, best predicts patch precision; we build on this to design an alignment-guided patch validation workflow in Section~\ref{subsec:eng-implications}.
}
\vspace{0.5em}
\section{Discussion}
\label{discussion}

\subsection{Engineering Implications}
\label{subsec:eng-implications}
Our findings suggest concrete design guidance across five classes of debugging automation pipelines: how bug reports are structured, and how that structure could inform test generation, fault localization, program repair, and report authoring itself. This guidance follows directly from our empirical results; we have not evaluated whether applying it improves downstream automation outcomes in practice.

\noindent\textbf{Structured Extraction (RQ1).}
RQ1 turns free-form bug reports into \emph{programmable artifacts}, structured representations that automated tools can consume directly, a goal shared by prior work such as BEE~\cite{song2020bee} and ChatBR~\cite{bo2024chatbr}, though these operate on the report in isolation rather than as an input contract for downstream generation. Three levers follow. First, since most extraction discrepancies reflect fill-versus-abstain policy rather than semantic error (Finding~1), conservative extraction suits high-risk settings such as fault localization, while coverage-oriented extraction better scaffolds generative tasks such as test synthesis. Second, the complementary extraction profiles of GPT-4o and DeepSeek-Chat suggest concrete ensemble strategies (Finding~3): pipelines prioritizing scenario reconstruction should favor DeepSeek-Chat, while exception-centric workflows benefit from cross-validating with GPT-4o. Third, since consistency is field-dependent (Finding~4), tooling should apply literal-overlap checks for entity anchors but switch to embedding-based similarity for scenario and behavioral fields. Structured extraction thus serves as a normalization front-end that reduces narrative ambiguity and stabilizes cross-artifact reasoning, the foundation on which the remaining levers build.

\noindent\textbf{Test Generation (RQ2).}
Findings 5--9 translate directly into design decisions for test-from-report pipelines, in the same spirit as prior LLM-based reproduction systems such as LIBRO~\cite{kang2023large}, Otter~\cite{ahmed2025otter}, and Issue2Test~\cite{nashid2025issue2test}, which we operationalize as a concrete algorithm below. Three levers emerge. First, RQ1's structured representations serve as standardized input contracts (exception, APIs, inputs, steps, expected/actual), reducing generation variance compared to raw narratives. Second, since alignment is primarily semantic rather than lexical (Finding~7), selection objectives should prioritize embedding-based similarity and explicit behavioral anchors over surface overlap. Third, because out-of-the-box LLM judges exhibit optimism and weak human rank alignment (Finding~6), deterministic proxies are preferable: OpenAI embedding cosine offers the most consistent approximation across datasets and representations (Finding~8). Together, these results enable an alignment-guided workflow that (i)~normalizes reports into anchor-rich schemas, (ii)~generates candidate tests conditioned on these anchors, and (iii)~ranks candidates using embedding-based proxies for behavioral encoding augmented with anchor coverage checks. This ranking reflects how faithfully a candidate encodes the report's stated behavior, not a validated measure of fault-detection adequacy, which we did not evaluate.

Algorithm~\ref{alg:testgen_alignment} operationalizes this workflow. Given a bug report $R$, we extract structured anchors $S$ (line~1, RQ1), reinforcing missing behavioral cues if anchors are sparse (line~3, Finding~9). Candidate tests are then generated explicitly conditioned on $S$: API and exception fields guide setup, input and step fields guide execution flow, and expected/actual fields guide oracle construction (Finding~7). Candidates are ranked using a deterministic behavioral-encoding score combining embedding cosine with anchor-coverage checks (line~12, Finding~8), normalized to $[0,1]$. This blueprint requires no alignment-specific model training and can be implemented with off-the-shelf LLM generators and embedding models.

\vspace{0.5em}
\begin{algorithm}[ht]
\vspace{0.5em}
\caption{Alignment-guided anchor-conditioned test generation}
\label{alg:testgen_alignment}
\footnotesize
\begin{algorithmic}[1]
\Require Bug report $R$ (summary+description), generator $G$, embedding model $E$, $N$ candidates, top-$k$
\Ensure Ranked tests $\mathcal{T}$ with behavioral-encoding scores

\State $S \gets \textsc{ExtractStructured}(R)$
\Comment{exception, API, inputs, steps, expected, actual --- RQ1}

\State $a \gets \textsc{AnchorScore}(S)$
\If{$a < \tau_a$}
  \State $S \gets \textsc{CompleteMissingAnchors}(R, S)$
  \Comment{reinforce missing APIs/steps/oracles --- Finding 9}
\EndIf

\State $\mathcal{C} \gets \emptyset$
\For{$i \gets 1$ to $N$}

  \State $c_i \gets \textsc{BuildConstraints}(S)$
  \Comment{API setup, reproduction steps, explicit oracle}

  \State $t_i \gets G(S, c_i)$
  \Comment{anchor-conditioned generation}

  \State $r_S \gets \textsc{ConcatFields}(S)$
  \State $r_{t_i} \gets \textsc{ReprTest}(t_i)$

  \State $s^{\text{embed}}_i \gets \cos\!\big(E(r_S),\,E(r_{t_i})\big)$
  \Comment{deterministic proxy --- Finding 8}
  \State $s^{\text{anch}}_i \gets \textsc{AnchorMatch}(S, t_i)$
  \State $s^{\text{oracle}}_i \gets \textsc{OracleCheck}(t_i)$

  \State $\hat{s}^{\text{embed}}_i, \hat{s}^{\text{anch}}_i \gets \textsc{Normalize}(\cdot)$

  \State $s_i \gets \alpha \hat{s}^{\text{embed}}_i
                  + \beta \hat{s}^{\text{anch}}_i
                  - \gamma (1 - s^{\text{oracle}}_i)$

  \State $\mathcal{C} \gets \mathcal{C} \cup \{(t_i, s_i)\}$

\EndFor

\State $\mathcal{C} \gets \textsc{Filter}(\mathcal{C}, \tau_s)$
\State $\mathcal{T} \gets \textsc{TopK}(\mathcal{C}, k)$
\State \Return $\mathcal{T}$

\end{algorithmic}
\end{algorithm}
\vspace{3mm}

\noindent\textbf{Fault Localization (RQ1--RQ2).}
Use behavioral anchors to constrain the search space: structured cues (APIs, exceptions, scenario steps) provide stable, near-canonical semantic queries for ranking suspicious components, since entity-level anchors show the highest cross-model extraction consistency of any field (Finding~4), extending prior IR-based localization work that relied on surface lexical similarity alone~\cite{zhou2012should,koyuncu2019d}. Prioritize embedding-based or code-aware matching over surface similarity, which can inflate with report length rather than with relevance (Findings~7, 9).

\noindent\textbf{Program Repair (RQ3).}
Findings 10--14 show that patch--report alignment is representation-sensitive and signal-dependent rather than intrinsic, complementing IR-and-fix-pattern-driven repair systems such as iFixR~\cite{koyuncu2019ifixr}. Three levers follow. First, evaluate patches as full diffs: contextual completeness provides the most stable alignment signal, while add-only or remove-only views lose semantic anchoring, especially under the STRUCTURED representation (Findings~10, 12). Second, prioritize embedding-based semantic alignment over surface overlap: code-aware embeddings (SBERT/CodeBERT/OpenAI) are the most stable signals across corpora (Finding~14). Third, condition generation and validation on report clarity: explicit expected/actual contrasts and a few salient steps strengthen alignment signals, whereas procedural length without contrast can dilute repair precision (Finding~13). Together, alignment can be operationalized as a controllable repair signal: (i)~normalize reports into structured behavioral schemas, (ii)~generate and evaluate patches as full diffs, and (iii)~rank candidates using embedding-based proximity augmented with anchor checks.

Algorithm~\ref{alg:patchgen_validate} mirrors the structure of Algorithm~\ref{alg:testgen_alignment}: anchors are extracted first, candidates are generated and scored via a deterministic proxy, and only the highest-scoring candidates are retained. We extract structured anchors $S$ and emphasize the expected/actual contrast (Finding~13), generate full-diff candidates by default (Finding~10), and score each via embedding cosine, anchor coverage, and a minimality prior that penalizes broad or unrelated edits, avoiding reliance on LLM judges, whose systematic optimism (Finding~11) makes them unsuitable for automated decision-making. Unlike test generation, patch validation has direct access to a ground-truth oracle: final validation applies the standard fail$\rightarrow$pass criterion against the failing test suite, optionally followed by regression tests.

\begin{algorithm}[ht]
\caption{Full-diff patch generation and validation guided by report anchors}
\label{alg:patchgen_validate}
\footnotesize
\begin{algorithmic}[1]
\Require Bug report $R$, codebase $C$, failing test suite $\mathcal{T}_{fail}$ (available), patch generator $G_P$, embedding model $E$,
$N$ candidates, top-$k$
\Ensure Validated patch(es) $\mathcal{P}$ and ranked candidates $\mathcal{C}$

\State $S \gets \textsc{ExtractStructured}(R)$
\Comment{expected/actual, exception, API, steps -- RQ1}
\State $S \gets \textsc{EmphasizeContrast}(S)$
\Comment{make expected vs actual explicit -- Finding 13}

\State $\mathcal{C} \gets \emptyset$
\For{$i \gets 1$ to $N$}
  \State $p_i \gets G_P(C,S)$
  \Comment{generate \emph{full diff} candidate by default -- Finding 10}
  \State $u^{embed}_i \gets \cos\!\big(E(\textsc{repr}(S)),\,E(\textsc{repr}(p_i))\big)$
  \Comment{deterministic proxy -- Finding 14}
  \State $u^{anch}_i \gets \textsc{AnchorMatch}(S,p_i)$
  \Comment{API/exception/entities/behavioral cues present}
  \State $u^{min}_i \gets \textsc{Minimality}(p_i)$
  \Comment{penalize broad edits / unrelated files}
  \State $u_i \gets \alpha u^{embed}_i + \beta u^{anch}_i + \lambda u^{min}_i$
  \State $\mathcal{C} \gets \mathcal{C} \cup \{(p_i,u_i)\}$
\EndFor

\State $\mathcal{C} \gets \textsc{TopK}(\mathcal{C},k)$
\State $\mathcal{P} \gets \emptyset$
\ForAll{$(p_i,u_i)\in \mathcal{C}$}
  \If{\textsc{Validate}$(C \oplus p_i,\mathcal{T}_{fail}) = \textsc{PASS}$}
    \State $\mathcal{P} \gets \mathcal{P} \cup \{p_i\}$
  \EndIf
\EndFor
\State \Return $(\mathcal{P},\mathcal{C})$
\end{algorithmic}
\vspace{0.5em}
\end{algorithm}

\vspace{0.5em}

\noindent\textbf{Bug Report Quality (RQ2--RQ3).}
Actionability depends on explicit anchors, not length (Finding~9). Naming APIs, specifying reproduction steps, and clearly stating expected vs.\ actual behavior improve downstream automation, whereas report length alone can dilute alignment despite increasing similarity scores (Findings~9, 13), reinforcing long-standing evidence that reproduction steps and an expected-versus-actual contrast are central to report usefulness~\cite{bettenburg2008makes,zimmermann2010makes}.
These five levers share a common principle: explicit behavioral anchors consistently outperform report length as a driver of automation quality, a pattern we revisit from the individual developer's perspective in Section~\ref{subsec:implication-developers}.

%\vspace{0.5em}
\subsection{Implications for Researchers}
\label{subsec:implication-researchers}

Our findings expose methodological blind spots in how empirical software engineering evaluates bug resolution, and suggest directions for more diagnostic, semantics-aware benchmarks.
\textbf{Behavioral alignment is not reducible to similarity.}
Across RQ2--RQ3, lexical overlap is minimal and embedding similarity only moderately reflects behavioral correspondence: similarity can inflate with report length, saturate with code-aware embeddings, or undervalue concise yet correct fixes. Researchers should separate \emph{surface resemblance} from \emph{behavioral correspondence} and explicitly measure the anchors linking reports to tests and patches.
\textbf{Structured representations materially change evaluation outcomes.}
An anchor-rich schema (API/exception, inputs, steps, expected vs.\ actual) reduces ambiguity and distinguishes extraction policy effects from genuine semantic errors; publishing and evaluating both the RAW and STRUCTURED representations improves comparability and diagnostic power.
\textbf{Patch view is a methodological variable.}
Full diffs provide the most stable alignment signal, while add-only and remove-only views shift both LLM judgments and metric behavior, so conclusions about report--patch alignment should be stratified by representation.
\textbf{LLM-based scoring is condition-sensitive rather than authoritative.}
Despite strong inter-human agreement, Human--LLM agreement remains modest and models exhibit systematic optimism that varies by representation and patch view; LLM scores should be reported with agreement and bias analyses rather than treated as ground truth.
Taken together, these results support a \emph{behavior-centric} framing of defect resolution as a propagation of semantic cues across artifacts, rather than a collection of isolated IR or APR tasks. Alignment provides a practical optimization signal for debugging automation: alignment scores can filter weakly grounded tests before they become repair oracles, rank candidate patches beyond fail-to-pass criteria, prioritize fault localization targets that preserve reported behavioral anchors, and detect under-specified bug reports lacking sufficient cues for reliable automation.

\subsection{Implications for Developers}
\label{subsec:implication-developers}

Our findings also translate into low-effort practices that make day-to-day LLM-assisted maintenance workflows more reliable, complementing the pipeline-level guidance of Section~\ref{subsec:eng-implications} with habits any developer can adopt individually.
\textbf{Write behavioral anchors into reports, not just narrative.} LLM tools behave more consistently when a report exposes a small set of checkable fields (API/exception, inputs, reproduction steps, expected vs.\ actual behavior) rather than a long free-form description, even without any tooling.
\textbf{Don't trust a passing test until it encodes the behavior.} A test that merely fails-then-passes is weak evidence: what matters is whether it \emph{explicitly encodes the reported behavior} (anchors + oracle). Before closing a bug, check that the test would fail for the \emph{right} reason.
\textbf{Review full diffs by default; use add/remove views only to probe intent.} Full diffs are the most reliable basis for judging whether a patch restores intended behavior. Reserve add-only or remove-only views for a specific question, such as whether a change merely suppresses the symptom, rather than using them as the primary review artifact.
\textbf{Treat an LLM's alignment or correctness score as a hint, not a verdict.} LLM judges are systematically optimistic and only weakly rank-aligned with human assessment. Cross-check any score against something concrete, such as whether the test covers the reported input condition or the patch addresses the expected/actual contrast, before acting on it.
Overall, stating behavioral anchors explicitly, insisting on tests with real oracles, and reviewing complete diffs are small habits with outsized effect: they are exactly the properties our results show LLM-assisted tools depend on to reason reliably about a bug.

\subsection{Threats to Validity}
\label{subsec:threats-to-validity}

% \noindent {\textbf{Threats to Internal Validity.}} 
%\noindent {\textbf{Construct validity.}}
\paragraph{Construct validity.}
\label{subsec:construct_validity}
Behavioral alignment between bug reports, tests, and patches is an abstract concept that we operationalize through three complementary proxies: structured anchors (the extracted exception/API/input/step/expected/actual fields), LLM-based judgments, and semantic similarity metrics. Each proxy captures only part of the phenomenon: lexical or embedding similarity may reflect surface overlap rather than true behavioral correspondence, a limitation consistent with prior critiques of embedding-based similarity as an imperfect proxy for developer intent~\cite{noyori2023deep,zhou2023leveraging}, while LLM judgments can be sensitive to framing, prompt design, or representation format. We mitigate this risk by triangulating across these three proxies, structured anchors, deterministic metrics, and LLM scores, and by validating our RQ2 and RQ3 alignment-score analyses against human references. The convergence of qualitative trends across these complementary measures reduces the likelihood that our findings are artifacts of any single operationalization.

\paragraph{Report provenance and completeness.}
\label{subsec:completeness}
A related threat concerns the provenance and completeness of the original bug reports themselves. Defects4J reports originate from a mix of developer-authored and user-authored issues. Developer-authored reports are sometimes filed primarily to track a fix the reporter has already identified, in which case the report may appear terse yet still translate into a well-aligned test and patch, not because the report itself is behaviorally rich, but because the reporter's downstream understanding was already complete at filing time. Conversely, user-authored reports may start incomplete and only converge toward a full behavioral understanding through subsequent discussion threads, with
the eventual test and patch reflecting this negotiated understanding rather than the original report text alone. Because our framework measures alignment between the \emph{initial} report and the artifacts eventually committed, this provenance distinction is a potential confound: it may inflate or deflate alignment scores independently of the intrinsic quality of the report's behavioral anchors. A related concern is how LLM judges respond to missing information: our weak correlations between individual field presence (e.g., \texttt{has\_api}, \texttt{has\_expected}) and judged alignment (Table~\ref{tab:rq2_llm_axis_top_attr}) are consistent with, though do not directly confirm, LLM scores not dropping substantially even when a behaviorally relevant field such as reproduction steps is absent from the report, which would be consistent with the broader optimism bias we
document elsewhere (Finding~1, Finding~6) rather than an accurate penalty for missing evidence. We do not currently have reliable per-bug labels distinguishing developer- from user-authored reports across our corpora, nor a targeted ablation isolating the effect of individual missing fields on judged alignment, and we did not control for either factor in the present study. We flag both as important directions for future work, and we encourage subsequent replications of Desc2Fix to stratify alignment analyses by report provenance and field completeness where such data is available.

%\noindent {\textbf{Internal validity.}}
\paragraph{Internal validity.}
\label{subsec:internal_validity}

Our results may be influenced by experimental design choices, including prompt wording, model configuration, and artifact representation. To assess robustness, we evaluate two distinct LLM families (GPT-4o and DeepSeek-Chat), two report representations (RAW and STRUCTURED), and three patch views (full, add, remove), and further analyze trends across Defects4J and SWT-Bench. While residual model-specific or prompt-specific effects cannot be entirely excluded, the stability of qualitative patterns across configurations and corpora supports the internal consistency of our conclusions.
A further internal validity concern relates to model and prompting recency. Our study evaluates GPT-4o and DeepSeek-Chat under a lightweight Chain-of-Thought strategy (Sec.~\ref{sub:prompting_techniques}), rather than native multi-step reasoning (``thinking'') modes or agentic orchestration frameworks that have since become more prevalent. This choice reflects the practical constraints of our experimental scope: two datasets, six report-representation/patch-view conditions, and four scoring dimensions, yielding a very large volume of alignment ratings that would be difficult to reproduce reliably and at comparable cost with reasoning-heavy or multi-agent pipelines. We note, however, that \textsc{Desc2Fix} evaluates alignment as a \emph{measurable signal over existing artifacts} rather than as a generation or repair capability to be maximized; the framework itself (structured anchors, deterministic proxies, and bias-aware LLM judging) is agnostic to which underlying model or prompting strategy produces the report--test--patch triplets or the judgments, and can be directly re-applied to newer reasoning-oriented or agentic models as they become available. We leave such replication to future work.

%\noindent {\textbf{External validity.}}
\paragraph{External validity.}
\label{subsec:external_validity}
We conduct our study on Defects4J and SWT-Bench, which span multiple projects and programming languages (Java and Python) and represent diverse bug categories. However, they do not cover all ecosystems (e.g., C/C++, mobile, industrial proprietary systems). Our experiments also involve two contemporary LLMs and widely used embedding models; future model generations or domain-specific systems may exhibit different behaviors. The consistency of observed patterns across these two datasets and artifact types suggests that the identified representation and alignment effects are unlikely to be artifacts of either single corpus; we cannot, however, claim that these effects generalize beyond Defects4J and SWT-Bench without further replication on additional ecosystems. Our structured schema and alignment framework are model-agnostic and designed to facilitate such replication on additional corpora.

%\noindent {\textbf{Reliability and annotation bias.}
\paragraph{Reliability and annotation bias.}
\label{subsec:reliability_and_annotation_bias}
Human evaluations were conducted by two annotators with software engineering expertise following task-specific guidelines and iterative calibration. We explicitly measured inter-annotator agreement using Spearman's $\rho$, Kendall's $\tau$, and MAE across evaluation dimensions. Agreement was consistently strong, particularly for correctness and specificity in RQ3, indicating stable and reproducible judgment criteria. All annotations were performed on the same fixed report sample, ensuring comparability across RQs and artifact types. Although manual assessment inherently involves some subjectivity, the high observed agreement supports the reliability of the human reference used throughout our analyses.

\section{Related Work}
\label{relatedwork}

\subsection{Bug Reports and Structured Information Extraction}
Bug reports are inherently noisy and heterogeneous, motivating extensive research on improving their structure and machine interpretability, from template-based report improvement~\cite{bettenburg2008makes} and IR/classification models~\cite{Lamkanfi2010PredictingTS} to fine-grained entity and relation extraction using NER and knowledge-aware representations~\cite{zhou2018recognizing,zhou2023leveraging}. More recently, trajectory-guided approaches such as TrajSpec~\cite{fahim2026bug} refine bug report specifications by mining repository execution trajectories to recover repair-relevant information reports often omit. These approaches enhance report understanding through entities, semantic relations, or trajectory-derived evidence, but treat bug reports as isolated textual artifacts (or, for trajectory-based refinement, enrich the report itself) without examining whether the resulting signals are preserved, transformed, or lost in downstream executable artifacts. In contrast, our work models structured behavioral anchors (exceptions, APIs, reproduction steps, expected vs. actual behavior) and evaluates their propagation across artifacts, moving beyond report enrichment toward semantic validation of the entire bug resolution pipeline.

\subsection{Bug Localization, Test Generation, and Report–Code Linking}
A substantial body of work investigates automated links between bug reports and code artifacts. Traditional fault localization combines spectrum-based and mutation-based signals~\cite{fraser2011evosuite}, more recent approaches integrate heterogeneous features into deep learning models such as DeepFL~\cite{li2019deepfl}, and IR-based methods remain common baselines~\cite{niu2025deep}, but predominantly rely on surface-level similarity. Test generation research has increasingly incorporated textual cues from bug reports, particularly in LLM-based settings, but typically evaluates generated tests by execution outcomes rather than explicit semantic correspondence with reported behavior, a gap made concrete by recent evidence that such tests can overfit to the observed failure without capturing the intended behavior~\cite{ahmed2026investigating}. Similarly, recent LLM-driven systems support patch generation, from single-pass pipelines~\cite{hossain2024deep,jin2023inferfix,zhang2404autocoderover} to autonomous agentic frameworks~\cite{bouzenia2025repairagent}, and test-generation work moves toward reproducing failures directly from issue text~\cite{ahmed2025otter,nashid2025issue2test}. A recent study finds that longer bug reports are associated with lower repair-agent resolution likelihood~\cite{khatib2026makes}, echoing our own observations on report length (Section~\ref{subsec:RQ2}, Finding 9) from an alignment-scoring rather than end-to-end repair-outcome perspective. Yet few studies examine whether generated or developer-written patches semantically reflect the report's behavioral signals. Our work instead treats report–test–patch relationships as a unified semantic propagation problem, quantifying how behavioral anchors flow across artifacts regardless of whether a stage is single-pass or agentic.

\subsection{Limitations of Embedding-Based Metrics for Semantic Alignment}
Vector-based similarity metrics (e.g., TF-IDF, SBERT, CodeBERT) are widely used to approximate semantic relatedness between bug reports and code artifacts~\cite{patil2023comparative}, but prior studies note that such metrics remain imperfect proxies for developer intent and behavioral nuance~\cite{noyori2023deep,zhou2023leveraging}. Our results empirically confirm this in a cross-artifact setting: embedding similarity can remain high for artifacts sharing entities or vocabulary while diverging in behavioral intent, and low for concise yet behaviorally correct tests or patches. Metric behavior is also representation-sensitive: the STRUCTURED representation reduces surface similarity while increasing predicate-level correspondence, and code-aware embeddings can approach ceiling values without guaranteeing behavioral restoration. We therefore advocate for behavior-aware evaluation frameworks that integrate structured signal extraction and cross-artifact validation rather than relying solely on vector similarity.

\subsection{Large Language Models and Semantic Validation}
LLMs have rapidly become central to automated program repair, code generation, and patch suggestion, with a recent systematic review of 189 studies highlighting the breadth of paradigms and evaluation strategies~\cite{zhang2024systematic}. The shift toward agentic repair pipelines motivates our focus on alignment as a foundational signal: a systematic study of six agentic and non-agentic repair systems on SWE-bench Verified~\cite{meng2024empirical} shows substantial performance variation and opaque failure modes, underscoring the need for artifact-level, model-agnostic alignment signals such as those \textsc{Desc2Fix} provides. LLMs have also been studied as evaluators of code and patch quality: Li et al.~\cite{li2025empirical} show that naturalness modeling can distinguish buggy, overfitting, and correctly repaired patches, and other approaches use self-evaluation such as round-trip validation~\cite{sharma2024patched}. Most existing protocols, however, rely on indirect proxies (compilation success, test-suite passing, token-level or distributional naturalness~\cite{sun2025empirical,yang2024revisiting}) that do not verify whether a patch restores the report's behavioral intent. A related study of LLM trust allocation across conflicting software artifacts finds that model confidence is poorly calibrated, with LLMs better at auditing natural-language specifications than detecting subtle code-level drift~\cite{ulfat2026measuring}, a calibration gap paralleling the optimism we observe in LLM-based alignment judgments (Sections~\ref{subsec:RQ2}--\ref{subsec:RQ3}). Our work complements this line of research by treating LLMs as behavioral alignment evaluators grounded in developer intent, integrating structured anchors, LLM-based scoring, deterministic metrics, and human-grounded validation into a reproducible cross-artifact framework, applicable to single-pass and agentic repair alike. To our knowledge, no prior work systematically quantifies behavioral signal propagation across reports, tests, and patches using both structured representations and human-calibrated LLM judgments.

\subsection{Issue Resolution and Reproduction}
A closely related line of work treats bug reports and GitHub issues as the starting point for reproducing and resolving failures end to end, often termed \emph{issue resolution} in the agentic and LLM-based literature. LIBRO~\cite{kang2023large} uses an LLM to generate and filter candidate failure-reproducing tests from a bug report, establishing reproduction as a task foundational to repair. Otter~\cite{ahmed2025otter} and Issue2Test~\cite{nashid2025issue2test} extend this to real-world GitHub issues. More recent work examines this pipeline critically: Ahmed et al.~\cite{ahmed2026investigating} show that tests generated for SWE-bench issues can overfit to the observed failure, echoing our own concerns (Section~\ref{subsec:RQ2}) that a passing test is not sufficient evidence of behavioral alignment; Fahim et al.~\cite{fahim2026bug} refine bug report specifications from execution trajectories, motivating our own discussion of report completeness (Section~\ref{subsec:completeness}); and Khatib et al.~\cite{khatib2026makes} identify which report-level features predict successful agent-based resolution. This body of work establishes issue reproduction as an active, scrutinized subfield, but evaluates success through execution outcomes rather than explicit, human-calibrated measurement of whether the report's behavioral intent is preserved across the resulting test and patch, the measurement gap \textsc{Desc2Fix} is designed to fill.

\section{Conclusion and Future Work}
\label{conclusion}
Bug resolution is not a single artifact task but a semantic propagation process: behavioral signals originate in natural-language reports and are expected to materialize in triggering tests and corrective patches, yet the extent of that propagation has remained largely unmeasured. In this work, we introduced \textsc{Desc2Fix}, a unified framework for quantifying how these signals are preserved, transformed, or lost across artifacts.
Our study demonstrates three core findings. First, structured behavioral anchors can be reliably extracted from bug reports, turning natural-language descriptions into stable and reproducible semantic input contracts. Second, cross-artifact alignment is measurable but highly representation-sensitive: lexical similarity alone is insufficient, patch view materially affects evaluation outcomes, and the STRUCTURED representation trades surface resemblance for stronger correspondence at the level of individual actions and entities. Third, LLM-based alignment judgments exhibit systematic optimism and only moderate agreement with humans, underscoring the need for bias-aware and multi-perspective evaluation.
Beyond empirical characterization, \textsc{Desc2Fix} reframes alignment as an actionable engineering signal rather than an observational property. Quantifying alignment could support early detection of weak or non-actionable reports, alignment-guided test generation, semantics-aware fault localization, and principled ranking of candidate patches in automated repair workflows, directions we sketch as reference algorithms in Section~\ref{subsec:eng-implications} but have not empirically validated. By grounding cross-artifact reasoning in structured anchors and reproducible metrics, our framework aims to strengthen the controllability and interpretability of LLM-driven maintenance pipelines, whether those pipelines are built from single-pass generators or agentic, tool-using systems.
\textbf{Future work} will extend this paradigm in four directions. First, we plan to incorporate additional artifact sources (e.g., stack traces, developer discussions) to model richer intent signals. Second, we plan to empirically validate the alignment-guided workflows sketched in Section~\ref{subsec:eng-implications}, testing whether alignment signals used as optimization objectives for test synthesis and patch generation actually improve downstream outcomes. Third, we envision interactive debugging workflows in which alignment feedback supports human--LLM collaboration, helping developers diagnose semantic drift and validate repair intent. Fourth, as reasoning-oriented and agentic models become more prevalent (Sec.~\ref{subsec:threats-to-validity}), we plan to re-apply \textsc{Desc2Fix} to such systems to test whether our representation- and patch-view-sensitivity findings generalize beyond single-pass instruction-tuned LLMs.
Ultimately, treating bug reports as controllable semantic drivers, rather than passive textual inputs, turns behavioral alignment from an afterthought of debugging automation into one of its central design variables.

\section*{Acknowledgements}{
This research was funded in whole, or in part, by the Luxembourg National Research Fund (FNR), grant reference AFR PhD bilateral, project reference 17185670. This work was also supported by the European Research Council (ERC) under the European Union’s Horizon 2020 research and innovation program (grant agreement No. 949014) and the Fundamental Research Funds for the Central Universities (AE89991/478). For the purpose of open access, and in fulfilment of the obligations arising from the grant agreement, the author has applied a Creative Commons Attribution 4.0 International (CC BY 4.0) license to any Author Accepted Manuscript version arising from this submission.
}

\bibliographystyle{ACM-Reference-Format}
\bibliography{references}

%%%Adding Appendix
\appendix

\section{Additional Tables}
\label{app:additional-tables}

\begin{table*}[ht]
\centering
\caption{Top bug-report attribute per deterministic metric (Spearman $\rho$), by dataset, scenario, and model.}
\label{tab:rq2_1_top_attr_per_metric}

\begin{subtable}{0.49\textwidth}
\centering
\caption{Defects4J}
\label{tab:rq2_1_top_attr_defects4j}
\scalebox{0.65}{
\begin{tabular}{@{}llllr@{}}
\toprule
\textbf{Scen.} & \textbf{Model} & \textbf{Metric} & \textbf{Best attribute} & $\boldsymbol{\rho}$ \\
\midrule

% -------- RAW (7 + 8 = 15) --------
\multirow{15}{*}{RAW}
  & \multirow{7}{*}{DeepSeek-Chat}
    & behavior\_sim       & num\_entities      & 0.160 \\
  & & codebert\_raw       & len\_report\_words & 0.339 \\
  & & entity\_sim         & num\_entities      & 0.231 \\
  & & rougeL\_raw         & avg\_field\_length & 0.188 \\
  & & sbert\_raw          & num\_entities      & 0.194 \\
  & & scenario\_sim       & has\_api           & 0.111 \\
  & & verb\_overlap\_raw  & len\_report\_words & 0.106 \\
\cmidrule(l){2-5}
  & \multirow{8}{*}{GPT-4o}
    & behavior\_sim       & has\_api           & 0.193 \\
  & & codebert\_raw       & has\_steps         & 0.367 \\
  & & entity\_sim         & has\_api           & 0.356 \\
  & & openai\_raw         & has\_api           & 0.239 \\
  & & rougeL\_raw         & has\_steps         & 0.245 \\
  & & sbert\_raw          & num\_entities      & 0.275 \\
  & & scenario\_sim       & has\_steps         & 0.434 \\
  & & verb\_overlap\_raw  & len\_report\_words & 0.124 \\

\midrule

\addlinespace[1.0mm]

% -------- STRUCTURED (4 + 5 = 9) --------
\multirow{9}{*}{STR.}
  & \multirow{4}{*}{DeepSeek-Chat}
    & codebert\_struct      & num\_entities      & 0.164 \\
  & & rougeL\_struct        & avg\_field\_length & 0.190 \\
  & & sbert\_struct         & num\_entities      & 0.162 \\
  & & verb\_overlap\_struct & avg\_field\_length & 0.159 \\
\cmidrule(l){2-5}
  & \multirow{5}{*}{GPT-4o}
    & codebert\_struct      & has\_steps         & 0.364 \\
  & & openai\_struct        & has\_api           & 0.246 \\
  & & rougeL\_struct        & avg\_field\_length & 0.287 \\
  & & sbert\_struct         & num\_entities      & 0.283 \\
  & & verb\_overlap\_struct & len\_report\_words & 0.175 \\

\bottomrule
\end{tabular}
}
\end{subtable}%
\hfill%
\begin{subtable}{0.49\textwidth}
\centering
\caption{SWT-Bench}
\label{tab:rq2_1_top_attr_swtbench}
\scalebox{0.70}{
\begin{tabular}{@{}llllr@{}}
\toprule
\textbf{Scen.} & \textbf{Model} & \textbf{Metric} & \textbf{Best attribute} & $\boldsymbol{\rho}$ \\
\midrule

% -------- RAW (10 + 8 = 18) --------
\multirow{18}{*}{RAW}
  & \multirow{10}{*}{DeepSeek-Chat}
    & behavior\_sim       & has\_api           & 0.078 \\
  & & codebert\_raw       & len\_report\_words & 0.223 \\
  & & entity\_sim         & num\_entities      & 0.197 \\
  & & jaccard\_raw        & len\_report\_words & 0.206 \\
  & & noun\_overlap\_raw  & has\_exception     & -0.074 \\
  & & openai\_raw         & len\_report\_words & 0.207 \\
  & & rougeL\_raw         & len\_report\_words & 0.217 \\
  & & sbert\_raw          & len\_report\_words & 0.287 \\
  & & scenario\_sim       & num\_entities      & 0.113 \\
  & & verb\_overlap\_raw  & len\_report\_words & 0.130 \\
\cmidrule(l){2-5}
  & \multirow{8}{*}{GPT-4o}
    & behavior\_sim       & has\_api           & 0.072 \\
  & & codebert\_raw       & num\_entities      & 0.262 \\
  & & entity\_sim         & has\_api           & 0.330 \\
  & & openai\_raw         & len\_report\_words & 0.215 \\
  & & rougeL\_raw         & len\_report\_words & 0.217 \\
  & & sbert\_raw          & len\_report\_words & 0.292 \\
  & & scenario\_sim       & has\_steps         & 0.388 \\
  & & verb\_overlap\_raw  & len\_report\_words & 0.139 \\
\midrule
\addlinespace[1.0mm]

% -------- STRUCTURED (7 + 5 = 12) --------
\multirow{12}{*}{STR.}
  & \multirow{7}{*}{DeepSeek-Chat}
    & codebert\_struct      & num\_entities      & 0.119 \\
  & & jaccard\_struct       & len\_report\_words & 0.139 \\
  & & noun\_overlap\_struct & has\_exception     & -0.082 \\
  & & openai\_struct        & avg\_field\_length & 0.176 \\
  & & rougeL\_struct        & len\_report\_words & 0.126 \\
  & & sbert\_struct         & len\_report\_words & 0.155 \\
  & & verb\_overlap\_struct & avg\_field\_length & 0.090 \\
\cmidrule(l){2-5}
  & \multirow{5}{*}{GPT-4o}
    & codebert\_struct      & num\_entities      & 0.215 \\
  & & openai\_struct        & has\_api           & 0.184 \\
  & & rougeL\_struct        & num\_entities      & 0.157 \\
  & & sbert\_struct         & num\_entities      & 0.211 \\
  & & verb\_overlap\_struct & len\_report\_words & 0.078 \\

\bottomrule
\end{tabular}
}
\end{subtable}

\begin{tablenotes}
\tiny
\centering
\item Scen.: scenario (RAW vs.\ STRUCTURED). STR.: structured.
\end{tablenotes}

\end{table*}

%%%%%%%%%%%%%%%%%%%%%%%%%%%%
\begin{table*}[t]
\centering
\caption{Patch-view effect on LLM scores (Wilcoxon signed-rank).}
\label{tab:rq3_wilcoxon_view_effect}

\scalebox{0.7}{
\begin{tabular}{@{}llllccc@{}}
\toprule
\textbf{Dataset} & \textbf{Representation} & \textbf{Model} & \textbf{Axis} &
\textbf{full--add} & \textbf{full--remove} & \textbf{add--remove} \\
\midrule

% ===================== Defects4J (16 rows) =====================
\multirow{16}{*}{Defects4J}
& \multirow{8}{*}{RAW}
  & \multirow{4}{*}{GPT-4o}
    & Alignment    & $-0.68^{***}$ (70)  & $-0.09$ (100)        & $+0.43^{***}$ (93) \\
& & & Coverage     & $-0.67^{***}$ (369) & $-0.28^{***}$ (492)  & $+0.45^{***}$ (352) \\
& & & Correctness  & $-0.58^{***}$ (369) & $-0.12^{**}$ (492)   & $+0.44^{***}$ (352) \\
& & & Specificity  & $-0.63^{***}$ (369) & $-0.25^{***}$ (492)  & $+0.41^{***}$ (352) \\
\cmidrule(l){3-7}
& & \multirow{4}{*}{DeepSeek-Chat}
    & Alignment    & $-0.17^{**}$ (369)  & $+0.23^{***}$ (492)  & $+0.37^{***}$ (356) \\
& & & Coverage     & $-0.25^{***}$ (370) & $+0.17^{***}$ (493)  & $+0.38^{***}$ (356) \\
& & & Correctness  & $-0.13^{*}$ (367)   & $+0.23^{***}$ (487)  & $+0.38^{***}$ (351) \\
& & & Specificity  & $-0.18^{***}$ (370) & $+0.15^{***}$ (493)  & $+0.32^{***}$ (356) \\
\cmidrule(l){2-7}

& \multirow{8}{*}{STRUCTURED}
  & \multirow{4}{*}{GPT-4o}
    & Alignment    & $-0.74^{***}$ (370) & $-0.51^{***}$ (494)  & $+0.24^{***}$ (355) \\
& & & Coverage     & $-0.73^{***}$ (370) & $-0.52^{***}$ (494)  & $+0.19^{***}$ (355) \\
& & & Correctness  & $-0.67^{***}$ (370) & $-0.43^{***}$ (494)  & $+0.17^{**}$ (355) \\
& & & Specificity  & $-0.66^{***}$ (370) & $-0.52^{***}$ (494)  & $+0.12^{*}$ (355) \\
\cmidrule(l){3-7}
& & \multirow{4}{*}{DeepSeek-Chat}
    & Alignment    & $-0.70^{***}$ (371) & $-0.17^{***}$ (494)  & $+0.39^{***}$ (356) \\
& & & Coverage     & $-0.68^{***}$ (371) & $-0.15^{***}$ (494)  & $+0.40^{***}$ (356) \\
& & & Correctness  & $-0.58^{***}$ (367) & $-0.10^{*}$ (494)    & $+0.30^{***}$ (352) \\
& & & Specificity  & $-0.71^{***}$ (371) & $-0.32^{***}$ (494)  & $+0.31^{***}$ (356) \\

\midrule

% ===================== SWTBench (16 rows) =====================
\multirow{16}{*}{SWT-Bench}
& \multirow{8}{*}{RAW}
  & \multirow{4}{*}{GPT-4o}
    & Alignment    & $-0.18^{***}$ (764)  & $-0.63^{***}$ (620)  & $-0.55^{***}$ (659) \\
& & & Coverage     & $-0.28^{***}$ (2321) & $-0.73^{***}$ (1971) & $-0.62^{***}$ (1956) \\
& & & Correctness  & $-0.30^{***}$ (2321) & $-0.75^{***}$ (1971) & $-0.66^{***}$ (1956) \\
& & & Specificity  & $-0.23^{***}$ (2321) & $-0.64^{***}$ (1971) & $-0.52^{***}$ (1956) \\
\cmidrule(l){3-7}
& & \multirow{4}{*}{DeepSeek-Chat}
    & Alignment    & $-0.26^{***}$ (2326) & $-0.32^{***}$ (1979) & $-0.11^{***}$ (1963) \\
& & & Coverage     & $-0.26^{***}$ (2326) & $-0.33^{***}$ (1979) & $-0.13^{***}$ (1963) \\
& & & Correctness  & $-0.26^{***}$ (2302) & $-0.33^{***}$ (1963) & $-0.12^{***}$ (1954) \\
& & & Specificity  & $-0.21^{***}$ (2326) & $-0.24^{***}$ (1979) & $-0.07^{**}$ (1963) \\
\cmidrule(l){2-7}

& \multirow{8}{*}{STRUCTURED}
  & \multirow{4}{*}{GPT-4o}
    & Alignment    & $-0.44^{***}$ (2321) & $-0.81^{***}$ (1973) & $-0.70^{***}$ (1958) \\
& & & Coverage     & $-0.39^{***}$ (2321) & $-0.81^{***}$ (1973) & $-0.72^{***}$ (1958) \\
& & & Correctness  & $-0.38^{***}$ (2321) & $-0.79^{***}$ (1973) & $-0.70^{***}$ (1958) \\
& & & Specificity  & $-0.27^{***}$ (2321) & $-0.71^{***}$ (1973) & $-0.57^{***}$ (1958) \\
\cmidrule(l){3-7}
& & \multirow{4}{*}{DeepSeek-Chat}
    & Alignment    & $-0.43^{***}$ (2326) & $-0.70^{***}$ (1979) & $-0.50^{***}$ (1963) \\
& & & Coverage     & $-0.40^{***}$ (2326) & $-0.69^{***}$ (1979) & $-0.50^{***}$ (1963) \\
& & & Correctness  & $-0.37^{***}$ (2308) & $-0.69^{***}$ (1957) & $-0.52^{***}$ (1945) \\
& & & Specificity  & $-0.36^{***}$ (2326) & $-0.66^{***}$ (1978) & $-0.45^{***}$ (1962) \\

\bottomrule
\end{tabular}
}

\vspace{2mm}
{\footnotesize \textit{Significance:} $^{***}p<0.001$, $^{**}p<0.01$, $^{*}p<0.05$; otherwise not significant.
Cells report $r$ with significance stars and paired sample size $n$ in parentheses.}
\end{table*}

%%%%%%%%%%%%
%%%%%%%%%%%%%%%%%%%%%%%%%%%%%%%%%%%%%%%%%%%%%
\begin{table}[t]
\centering
\caption{Model effect on LLM scores (Wilcoxon signed-rank).}
\label{tab:wilcoxon_model_effect_fullcorpus}

\scalebox{0.7}{
\begin{tabular}{@{}llcccc@{}}
\toprule
\textbf{Dataset} & \textbf{Scenario} & \textbf{Alignment} & \textbf{Coverage} & \textbf{Correctness} & \textbf{Specificity} \\
\midrule

\multirow{2}{*}{Defects4J}
& RAW        & $+0.46^{***}$ (501) & $+0.23^{***}$ (501) & $+0.48^{***}$ (501) & $+0.41^{***}$ (501) \\
& STRUCTURED & $-0.13^{**}$ (502)  & $-0.10^{*}$ (502)   & $-0.14^{**}$ (502)  & $-0.10^{*}$ (502) \\
\midrule

\multirow{2}{*}{SWT-Bench}
& RAW        & $+0.43^{***}$ (2344) & $+0.20^{***}$ (2344) & $+0.58^{***}$ (2342) & $+0.37^{***}$ (2344) \\
& STRUCTURED & $-0.04^{*}$ (2340)   & $+0.17^{***}$ (2340) & $+0.04$ (2339)       & $-0.12^{***}$ (2340) \\
\bottomrule
\end{tabular}
}

\vspace{1.5mm}
{\footnotesize \textit{Significance:} $^{***}p<0.001$, $^{**}p<0.01$, $^{*}p<0.05$; otherwise not significant.
Cells report effect size $r$ with significance stars and paired sample size $n$ in parentheses.}
\end{table}

\end{document}